%% file: paper.tex
%**start of header
\input newhead3

\input xdefs3.tex
\input layout.tex

\input table.tex
\def\GG{{\cal G}}
\def\KK{{\cal K}}
\def\ie{{\it i.e.}}
\def\kout{{k-{\rm out}}}
\def\kreg{{k-{\rm reg}}}
\def\text#1{\leavevmode\hbox{#1}}

\let\epsilon=\varepsilon
\let\truett=\tt
\fontdimen3\tentt=2pt\fontdimen4\tentt=2pt
\def\tt{\hfill\break\null\kern -2truecm\truett ************ }
\def\ka{k}
\def\kap{\ka}
\def\kb{2k-2}
\def\kbp{(\kb)}
\def\kc{k-2}
\def\kcp{(\kc)}
\def\factor{2k(5k+1)}
\def\ltr{\ell_{\rm triang}}
\newcount\REMCOUNT
\let\rho=\varrho
\REMCOUNT=0
%\DRAFT
\def\REMARK#1#2{
\vskip.1in\medbreak\noindent%
{\lmargin{\the\REMCOUNT}\bf Remark\ \actualnumber.\number\CLAIMcount:
}{\sl #2}%
\NEWDEF{r}{#1}{Remark\ \actualnumber.\number\CLAIMcount}%
\global\advance\CLAIMcount by 1%
\global\advance\REMCOUNT by 1%
\ifdim\lastskip<\medskipamount%
\removelastskip\penalty55\medskip\fi
}
%\input /users/eckmann/papers/tools/xdefs
%**end of header
{\titlefont{\centerline{The Number of Large Graphs }}}
\vskip 0.5truecm
{\titlefont{\centerline{{with a Positive Density of Triangles}}}}
\vskip 0.5truecm
{\it{\centerline{ P. Collet${}^{1}$ and J.-P. Eckmann${}^{2,3}$}}}
\vskip 0.3truecm
{\eightpoint
\centerline{${}^1$Centre de Physique Th\'eorique, Laboratoire CNRS UMR
7644,
Ecole Polytechnique, F-91128 Palaiseau Cedex, France}
\centerline{${}^2$D\'ept.~de Physique Th\'eorique, Universit\'e de Gen\`eve,
CH-1211 Gen\`eve 4, Switzerland}
\centerline{${}^3$Section de Math\'ematiques, Universit\'e de Gen\`eve,
CH-1211 Gen\`eve 4, Switzerland}
}
\vskip 0.5truecm\headline
{\ifnum\pageno>1 {\toplinefont Graphs with a Density of Triangles}
\hfill{\pagenumberfont\folio}\fi}
\vskip 1cm
{\eightpoint\narrower\baselineskip 11pt
\LIKEREMARK{Abstract}We give upper and lower bounds on the number of
graphs of fixed degree which have a positive density of
triangles. In particular, we show that there are very few such graphs, 
when compared to the number of graphs without this restriction.
We also show that in this case the triangles seem to
cluster even at low density.
}
\SECTION{Introduction and statement of results}In a number of contexts
involving large graphs (such as the Web or Citation Networks)
it has been observed that such graphs contain a large number of
triangles,
and probably a positive density of them (per node). We refer to [EM]
for a detailed discussion. The importance of topology is also
mentioned in [AB], where the authors say (p.41):

{\narrower{\it\smallskip\noindent 
``But if the topology of
these networks indeed deviates from a random graph, we need to develop
tools and measurements to capture in quantitative terms the underlying
organizing principles.''}\smallskip}

On the other hand, it is
 well known that in several models of random graphs with a bounded
number of links per node
the probability of
observing a large number of triangles is vanishingly small when the number 
of nodes diverges. A natural question is therefore to estimate more precisely 
the number of graphs with a large number  of triangles. It will become
clear from the discussion of our paper that this result is beyond the
``large deviation bounds'' which are found in the literature [VV, JLR].

In this paper we study
the cardinality of sets of graphs with a {\em positive density} of
triangles per node. We consider (random) graphs with sparse sets of
links,
\ie,
random graphs in which the number of links is bounded by a fixed constant
times the number of nodes. We will consider three models of labeled graphs:
\setitemindent{G)}
\item{G)}The model
$\GG_{n,k}$ comprises the graphs with $n$ nodes and $kn$ links. We
call them $k$-general graphs.
\item{O)}The model
$\GG_{n,\kout}$ is the set of all graphs with $n$ nodes, and from each
node there leave exactly $k$ directed links (directed from that
node). We call these graphs $k$-out.
\item{R)}The model $\GG_{n,\kreg}$ is the set of all graphs where at
each node exactly $k$ links meet. (This definition is only interesting
if $kn$ is even, which we tacitly assume in the sequel.) These graphs
are called $k$-regular.

A well-studied question is that of the probability of finding
triangles in such graphs, where the probability is relative to the
uniform measure on the set of graphs, giving the same weight to each
graph. For all of the above examples, it is known that the expected
number of triangles in these graphs is {\em bounded} independently of
$n$, by a quantity $\lambda =\OO(k^3)$. Furthermore, for each $t\ge1$,
it has been shown that the probability to find exactly $t$ triangles
is given, in the limit $n\to\infty $, by the Poisson distribution
$$
P(t)=e^{-\lambda }{\lambda ^t\over t!}~.
$$
Note however, that this limit is not at all uniform in $t$, as will be
illustrated by our results in Sec.~4. Further studies have greatly
refined this result, giving very precise estimates on the tails of
this distribution, as a sort of large deviation result.
Our study, in this paper, goes beyond that region, since we ask for
the size of subsets of the three graph families with a
{\bf positive  density} of
triangles. We assume throughout that $\alpha $ is a fixed constant
$\alpha >0$ and we consider those graphs in the above classes which
have $\alpha n$ triangles (or, more precisely $[\alpha n]$ triangles,
where $[x]$ denotes the integer part of $x$).
We denote these subsets by $\GG_{n,k,\alpha }$, $\GG_{n,\kout,\alpha
}$, $\GG_{n,\kreg,\alpha }$.
If $X$ is a finite set we denote by $|X|$ its cardinality.
Our main result is the
following
\CLAIM{Theorem}{all}Fix $k\in \natural$ and $\alpha >0$.
For the three graph families we have the bounds (valid when
the lower bound is non-negative):
$$
\eqalignno{
&\hphantom{\,\le,}\lim_{n\to\infty } {\log|\GG_{n,k,\alpha }|\over\log|\GG_{n,k
}|}\,=\,1~,\NR{11}
1-{3\alpha \over4( k^2-1)}\,&\le\,\liminf_{n\to\infty }
{\log|\GG_{n,\kout,\alpha }|\over \log|\GG_{n,\kout
}|}\cr\,&\le\,\limsup_{n\to\infty } {\log|\GG_{n,\kout,\alpha }|\over
\log|\GG_{n,\kout
}|}\,\le\,1-{\alpha \over 2k^2(5k+1)}~,~~\NR{2}
1-{12\alpha \over k^2-1}&\,\le\,\liminf_{n\to\infty } {\log|\GG_{n,\kreg,\alpha
}|\over \log|\GG_{n,\kreg
}|}\cr\,&\le\,\limsup_{n\to\infty } {\log|\GG_{n,\kreg,\alpha }|\over
\log|\GG_{n,\kreg
}|}\,\le\,1-{2\alpha \over k(k-1)}~.~~\NR{3}
}
$$

We conjecture that in the statement above the limits exist (assuming,
of course, that $kn$ is even in the $k$-regular case).

\REMARK{000}{From the point of view of Information Theory or Statistical
Mechanics/Large Deviations, the number of triangles is an extensive
quantity relative to the number of nodes.
But the logarithmic bounds we find are
{\bf not}
extensive in the number of nodes:
They are extensive and small on the scale (of the logarithms) of the number of graphs.
This suggests that the presence of a
positive density of triangles is a very strong information about the
system.\hfill\break\indent
Indeed, imposing that the number of triangles is proportional to the
number of nodes leads intuitively to the conclusion that if one
considers two links emanating from a common node, there is a non zero
probability that their ends are also linked.}

\REMARK{1}{We prove more precise bounds in \equ{fn}.}
\REMARK{2}{One should note that a $k$-regular graph is more like a
$k/2$-out graph (because each link is counted twice).}
\REMARK{2a}{The lower bounds are obtained by constructing graphs
containing complete graphs of maximal size. We do not know whether
these bounds are optimal. If they are, this would mean
that complete graphs are ``typical'' among random graphs with a
positive density of triangles.}

The reader should observe that the lower bound is a little
surprising. Indeed, assume $\alpha >0$ is very close to 0.
Then, one might expect that since the density of triangles is very
low, they will typically be (edge and node) disjoint in the
set $\GG_{n,\kout,\alpha 
}$ (and similarly for $\GG_{n,\kreg,\alpha}$). 
For each such triangle, once one has placed 2 of the 3 links forming
it, the third link is already determined when we close the 
triangle
and thus, under the assumption of disjointness,
only ${nk-\alpha n}$ links can be chosen freely [in the case of the
regular graphs this number is ${nk/2-\alpha n}$] leading to an upper bound
of $n^{nk-\alpha n}$. But, our lower bound is larger than
that. Therefore we conclude: 
\REMARK{cl}{In the families of graphs
$\GG_{n,\kout,\alpha  
}$ and $\GG_{n,\kreg,\alpha }$ the triangles have a natural tendency
to coagulate into clusters.}

Another surprising result is that in the case of $\GG_{k,n,\alpha }$ of
Eq.\equ{11}, there
is {\em no loss} in the number of graphs on the scale of
$n^n$. This might seem all the more surprising in view of the Poisson
distribution of the expected number of triangles (but is more credible
in view of the
non-uniform diagonal limit $t=\alpha n\to\infty $ which we are considering).

The paper proceeds from the $k$-regular graphs via the $k$-out graphs
to the $k$-general graphs. We prove first upper bounds and then the
(easier) lower bounds.

\SECTION{Upper bound for $k$-regular graphs}This
section should be considered as a warm-up for the next
one. Therefore, many arguments are sketched, and the reader can find
longer explanations in the next section. On the other hand, the
general line of proof should be more transparent. The reader will also
notice that the $k$-regular case is much less delicate than the
$k$-out case.

We assume that $nk$ is
even because otherwise there are no $k$-regular
graphs.
For a $k$-regular graph,
the general bound is [B]:
$$
|\GG_{n,\kreg}|\,\approx\, C(k)^n n^{nk/2}~.
\EQ{reg0}
$$
A given link cannot be an edge in more than $k-1$ triangles, because
exactly $k$ links meet at each node (see also
\clm{abc}).
For every
link we say that it is $s$ times occupied if it occurs in $s$
triangles, and every triangle occupies (in this sense) 3 links. Thus
the total occupation number is $3\alpha n$, and therefore the number
of links involved in edges of triangles is at least $3\alpha n/(k-1)$ (and
at most $3\alpha n$). We next bound the number of ways to draw $\alpha
n$ triangles. Label the nodes, and for every node $i$
let $t_i$ be the number of triangles having $i,j,m$ as corners with
$i$ the {\em smallest} of the three indices. The number of ways to
choose the $t_i$ is
$$
{\alpha n +n-1  \choose n-1}\,\le\, 2^{(\alpha +1)n}~,
\EQ{reg1}
$$
which is negligible on the scale we consider.
With the $t_i$ fixed, we draw the
triangles at $i=1,2,\dots$. {\em Note that in this process we will
have to place at least $\ell=3\alpha n/(k-1)$ links.}
Now, if we draw a triangle, several things can happen. Either the
triangle is already drawn, because its 3 sides have been placed as
sides of triangles which have been drawn earlier. No link needs to be
placed in this case.
In all other cases, between one and 3 links need to be drawn.
The least favorable case occurs when 3 links have to be placed.
Then, there are at most $n(n-1)$ possibilities to choose the first 2 links and
then at most 1 possibility for the third, and we get a factor of $n^{1-(1/3)}$
per
link.
If only one new link is used, and it starts at $i$ (and the two others are
already there), there are at most $k^2$ possible endpoints for that
link and we get a factor $k^2n^{1-1}$ in this case. If the one
missing link is between two links (which are already there when this
link has to be placed) we get at most $k(k-1) n^0 /2$
possibilities. Finally, if two links are missing, there are 2
possibilities: Either it is two links starting at $i$, and this makes
at most $nk=kn^{2-1}$ ways, or one link starting at $i$ and one not starting
at $i$ which makes at most $n k=n^{2-1}k$ possibilities. Indeed, there
are $n$ possibilities to 
choose the end $j$ of the link starting at $i$
and then there are at most $k$ possibilities for choosing the second link.
Once two links are chosen, the triangle is completely determined since
all its nodes are fixed.

Thus for all these links we get a bound of at most
$$
k^{2\ell} n^{\ell -\ell/3}
\EQ{reg3}
$$
possibilities, that is, $k^2 n^{2/3}$ {\em per link}.
Finally, the remaining $kn/2-\ell$ links can be put in at most
$n^{kn/2-\ell} $ ways. Summing over the possible number of links
(which is bounded by $3\alpha n/(k-1)\le\ell\le3\alpha n$ and yields a
factor which can be easily absorbed), and
combining the two bounds, we get a bound
$$
C^n n^{kn/2-3\alpha n/(3(k-1))}\,=\,
C^n n^{kn/2-\alpha n/(k-1)}~.
\EQ{reg4}
$$
This completes the proof of the upper bound of \equ{3}.\QED
\LIKEREMARK{Remark}A second proof could be derived from a
modification of the proof for the case of $k$-out graphs which we give
below.

\SECTION{An upper bound for $k$-out graphs}In this section, we
consider the set
$\GG_{n,\kout}$ of graphs where each node has $k$ out-links.
The cardinality of this set is
$$
|\GG_{n,\kout}|\,=\, {n-1 \choose k}^n~.
\EQ{1}
$$
In other words, we allow for links which go back and forth between 2
nodes, but we do not allow double directed links in the same direction
between 2 nodes. Also self-links (loops) are forbidden.
We denote by $\GG_{n,\kout,\alpha }$ the subset of $\GG_{n,\kout}$ with
$[\alpha
n]$ triangles, where triangles are counted as follows: Once the links
are placed, their orientation is neglected and unoriented triangles
are counted, including the multiplicity of the edges (which can be 1
or 2 by what we said above). For example, 3 nodes with the possible 6
directed links between them count as $8=2^3$ triangles. By and large,
these distinctions are not very essential for the proofs we are going
to give and other choices will work with similar proofs.

Since, for fixed $j$, one has
$$
{(m-j+1)^j\over j!}\,\le\,{m\choose j}\,\le\,{m^j\over j!}~,
$$
we see that, for fixed $k$,
$$
\lim _{n\to \infty}{\log |\GG_{n,\kout}|\over n\log n}\,=\, k~,
$$
which we will sometimes write in the more suggestive form
$$
|\GG_{n,\kout}|\,\approx\, n^{nk}~.
$$
To be more precise, we define the notation $F(n)\,\approx\, n^{nc}$ to
mean that there are constants $C_1>0$ and $C_2$ independent of $n$
(but not of $k$) such that
$$
C_1^n n^{nc}\,\le\,F(n)\,\le\, C_2^n n^{nc}~,
\EQ{fn}
$$
so that the error term in the limit is subexponential. Another way to say
this is
$$
\log |\GG_{n,\kout}| \,=\, n\bigl(k\log n +\OO(1)\bigr )~.
$$
Define
$$
\rho(k)\,=\,{1 \over \factor}~.
$$
\CLAIM{Proposition}{upper}{There is a $C=C(\alpha ,k)<\infty $ for which
the quantity $|\GG_{n,\kout,\alpha }|$ satisfies
an upper bound of the form
$$
|\GG_{n,\kout,\alpha }|\,\le\, C^n n^{n(k -\alpha \rho (k))}~.
\EQ{upper}
$$
}

\REMARK{3}{This is the upper bound of \equ{2}.}
To avoid the notation $[\alpha n]$, we
assume henceforth that $\alpha n$ is an integer. We consider a
configuration with $\alpha n$ triangles. The triangles which can occur
in a $k$-out graph are of two
types, which we call type $R$ (for round) and type $F$ for (for
frustrated) depending on the relative orientation of the links.
\figurewithtex{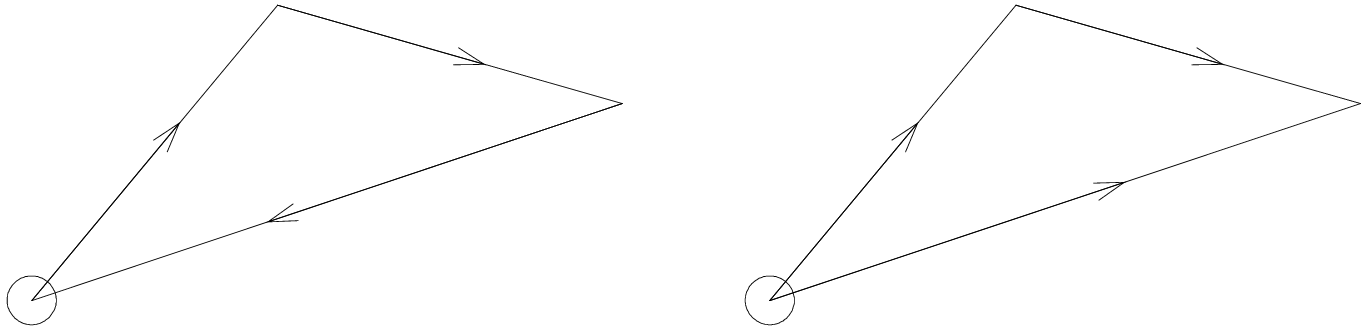}{triang.tex}{3.5}{15}{``Round'' and
``frustrated'' triangles. In the first case all links ``follow each
other'' while in the second there is a ``reverse'' (frustrated) link,
the link $c$. The corner with a circle is called the anchor of the
triangle, and the links are then labeled in such a way that for a
round triangle the $a$ link leaves the anchor, and the others follow
in order, while for the frustrated triangles, the $a$ link leaves the
anchor and the $b$ link leaves the end of the $a$ link. These rules
determine a unique labeling of each triangle if we require the anchor
for the round triangle to be at the node with lowest number.}\cr

We next consider the
number of triangles in which a given edge can occur.
Because of the $k$-out model, edges of type $b$ can occur in arbitrary
many triangles of type $F$, by just connecting 2 lines from any node
to a given edge. In this respect, the $k$-out model is more
complicated than the $k$-regular model.
However, the other lines can occur only in a small
number of triangles.
\setitemindent{--}
\CLAIM{Lemma}{abc}Bounds on the number triangles per link:
\item{--}A link can be an edge of type $a$ in at most $k$
triangles of type $R$ and in at most $k-1$ triangles of type $F$.
\item{--}A link can be an edge of type $b$ in at most $k$
triangles of type $R$.
\item{--}A link can be an edge of type $c$ in at most $k$
triangles of type $R$ and in at most $k-1$ triangles of type $F$.

\PROOF Consider first the case $R$. The edge $a$ can
occur in at most $k$ triangles. To see this, note that once $a$ is
placed, there are $k$ edges of type $b$ leaving its end, and then
the triangle must be closed, so there are at most $k$ such
triangles. Since $R$ is round, the same reasoning can be done for
the other edges.
In the case of a triangle of type $F$, we have already remarked that there
is no bound possible for the link $b$, but we claim the others cannot
be part of more than $k-1$ triangles of type $F$. Indeed, once link
$a$ is fixed, we need to choose another out-link to become link $c$
(and then the  $b$ link is fixed). This gives $k-1$ as a bound.
Finally, link $c$ can belong only to $k-1$ triangles of type $F$
because for fixed $c$ there remain only $k-1$ candidates for
the edge of type $a$.\QED

An important consequence of \clm{abc} is that the
number of edges belonging to at least one of the $\alpha n$ triangles grows
proportionally with $\alpha n$:
\CLAIM{Lemma}{triangles}The number of edges $\ltr$
belonging to at least one of the $\alpha n$
triangles in a graph of type $\GG_{n,\kout}$ is bounded by
$$
{\alpha n \over 2k}\,\le\,\ltr\,\le\,3\alpha n~.
\EQ{triang}
$$

\PROOF The upper bound is obvious. To prove the lower bound, note that
every triangle involves a link of type $a$. Since there are $k$ links
leaving from the far end of that link, the number of triangles for
which this link is an $a$ link is bounded above by $2k$ ($k$ of type
$F$ and $k$ of type $R$). Thus, at least
${\alpha n\over 2k}$ links are needed just to draw all $a$ links.\QED
\REMARK{4}{Note that an $a$ link can be also a $b$ or $c$ link for many
other triangles, and so the above argument cannot be easily
improved. When $k=2$ and $n=3$ the complete graph forms 8 triangles,
but needs 6 links, 
instead of the 4 as given by the lower bound. For complete, directed,
$k$-out graphs the asymptotic bound for $k\to \infty $ is ${3\alpha
n\over 4k}$.}
To prove the bound of \clm{upper} we give an algorithm which constructs
all the graphs with $\alpha n$ triangles, and perhaps a few more with
more triangles, and we bound the number of ways in which this can be
done.

To enumerate all the cases, we first label the nodes in an arbitrary
fashion from 1 to $n$. Once this is done, we consider any
configuration with $\alpha n$ triangles. We associate each triangle
with a node as follows: Triangles of type $F$ are associated with the
node from which the $a$ and $c$ links originate. For triangles of type
$R$ we label the edges in such a way that the corner where the $a$ and
the $c$ edges meet has the lowest label among the 3 corners.
We call the point from which the $a$ link leaves the {\em anchor} of
the triangle.

Once this is done, there will be $t_i$ triangles anchored at node
$i$, for $i=1,\dots,n$.
Furthermore, we denote by $v_i$ the number of links arriving at node
$i$ {\em once the graph will have been completely constructed}.
Both $t_i$ and $v_i$ indicate the values {\em at the end} of
constructing the graph.

In order to construct all possible graphs with $\alpha n$ triangles,
we start by choosing the $t_i$ and the $v_i$.
Clearly,
$$
\sum_{i=1}^n t_i\,=\,\alpha n~,
\EQ{t1}
$$
and therefore the  number of ways of choosing the $t_i$ is bounded by
$$
A\,=\,{ \alpha n +(n-1)\choose n-1}\,\le\,2^{(\alpha+1)n} ~.
\EQ{A}
$$
The $v_i$ satisfy $\sum_i v_i= kn$, since each link arrives
somewhere. The number of ways to distribute the ends of the $kn$ links
is
therefore bounded above by
$$
B\,=\,{kn + (n-1)\choose n-1}\,\le\, 2^{(k+1)n}~.
\EQ{B}
$$

We will need the following
\CLAIM{Lemma}{vi}The product of the $v_i+k$ satisfies
$$
\prod_{i=1}^n (v_i+k)\,\le\, (2k)^n~.
\EQ{vi}
$$

\PROOF Since the sum of the $v_i+k$ equals $2kn$, the maximal value
of the product is $\bigl(2kn/n\bigr)^n$.\hfill\break
\hphantom{000}\QED

Thus, once the $t_i$ and $v_i$ are fixed, we have lost a
(negligible) combinatorial factor $C_0^n$, with $C_0\le 2^{\alpha +k+2}
(k+1)$.

\CLAIM{Lemma}{ti}At each node, at most
$k^2$ round triangles and at most $k(k-1)$ frustrated triangles can be
anchored. 

\PROOF Consider first the round triangles. There are $k$ outgoing
links from a given node, and from each of their ends there are another
$k$ outgoing links and then the triangle must be closed, and so there
are at most
$k^2$ round triangles. For the frustrated triangles, we first choose a pair of
outgoing links and then the direction of the link connecting their
ends. \QED

Having fixed the $t_i$, $v_i$,
we now place the triangles starting with all those anchored at node 1,
proceeding to node 2, 3, and so on, until we arrive at node $n$.
At each node $i$
we construct first all the $F$ triangles and then all the $R$
triangles.
Assume the first $s-1$ triangles have been drawn, and assume we are
placing the next triangle anchored at node $i$. 
We will first make a choice of which links of the new triangle are
assumed to be present. This gives $8=2^3$ choices. There are 2 more
choices between type $F$ and $R$, (in fact less since we insist on
building first all the $F$ before the $R$). For each of these
choices, we bound the maximal number of ways a triangle can be placed.
We call these bounds $F_j$ for the frustrated triangles and $R_j$
for 
the round triangles, $j=1,\dots,8$. An upper bound on the number of
ways to place 
a triangle (given its anchor) is then
$$
16 \max(F_1,\dots,F_8,R_1,\dots,R_8)~.
$$

When we construct a triangle at $i$, it will be denoted by its corners
$(i,j,m)$. If it is round its links are $a=ij$, $b=jm$,
and $c=pq$ with $p=m$, $q=i$.
If it is frustrated, its links are $a=ij$, $b=jm$,
and $c=pq$ with $p=i$, $q=m$.
\midinsert
\medskip
\centerline{\bf Table I}
\medskip
\begintable
~~case~~\|~new links~|~\# links$=\delta q$~|~~~$R$~~~|~~~$F$~~~|~~~~min.~gain~~~~\crthick
1\|$ abc    $|$3$|$n^2  $|$n^2$|$n^{-1} $\cr
2\|$ ab $|$2$|$n v_i$|$nk  $|$(v_i+k)n^{-1}$\cr
3\|$ ac    $|$2$|$nk$|$nk   $|$k n^{-1}$\cr
4\|$ a$|$1$|$n$|$n  $|$ 1 $ $\star$\cr
5\|$ bc    $|$2$|$nk$|$nk   $|$k n^{-1}$\cr
6\|$ b$|$1$|$k v_i $|$k^2 $|$(v_i+k)^2 n^{-1}$\cr
7\|$c      $|$1$|$k^2 $|$k^2  $| $k^2 n^{-1}$\cr
8\|none|$0$|$k^2$|$ k^2  $|$ k^2 $ $\star$
\endtable
\endinsert

The 16 cases are represented in Table I. The second column
indicates 
which links are new in forming the triangle, and the next the number
of these new links.
The next two columns indicate the maximum number of ways the given
case can appear.
The last column will be explained below.
\LIKEREMARK{Proof of Table I}To prove the bounds on the multiplicative
factors is
just a verification. We indicate a few cases to guide the reader. In
case 1, we place 3 links of which 2 can be chosen freely (the $a$
link and then the $b$ link), whereas the third link is then completely
determined. Therefore, we get a factor $(n-1)(n-2)$, for both types of
triangles, and we bound this by $n^2$ In case 2, the $c$ link is already present. For a round
triangle there are at most $v_i$ possibilities for a $c$ link to end in
$i$. The $a$ link can be chosen in $n$ ways, and the $b$ link must
connect the end of $a$ to one of the $c$, and this can be done in
$v_i$ ways. Thus we get a factor $nv_i$. In the case of the frustrated
triangles, there are only $k-1$ possibilities for the $c$ link (which
now {\em originates} at $i$) since one link is used as the $a$ link:
The factor is therefore at most $nk$. All other cases are discussed similarly,
for example, in case 7, the $c$ link is missing, but the $a$ and $b$
links are present, and there are $k$ possible ends for $a$ and another
$k$ possible ends for each of the $b$ attached to $a$.
Finally, we explain case 4 which is the critical case. In it,
the $b$ link and the $c$ link are given. Since one link cannot be
placed
in more than $n-1$ ways, the factor $n$ is an upper bound.
Finally, the last column of Table~I is calculated as follows. Its
entry is an upper bound on the sup of the $R$ and $F$ column, divided
by $n^{\rm\#~links}$.
 \QED
\smallskip

To explain the proof of \clm{upper} we will first consider the simpler
situation where the cases 4 and 8 do not appear. Indeed, in these two
cases, the combinatorial factor of the last column in Table I is {\em
not} small when $n$ is large. 
In this simplified case, each time we place a triangle, the number of
links increases by $\delta q$ and the number of possibilities is
bounded by 
$$
16 n^{\delta q} n^{-1} (v_i+k)^2~.
$$
Since $1\le\delta q\le 3$ we can bound this from above by
$$
16 n^{\delta q (1-1/3)} (v_i+k)^{2}~.
$$
From \clm{triangles}, we know limits on $\ltr$, so that
$\sum (\delta q)=\ltr$, and $\ltr\in[\alpha n/(2k),3\alpha n]$.
We get, in the end, for
constructing at least $\alpha n$ triangles with $\ltr$ links,
an upper bound of
$$
16^{\alpha n} n^{2\ltr/3} \prod_{i=1}^n (v_i+k)^{4k^2}~.
$$
The last factor is obtained by observing that there are at most
$k^2+k(k-1)$ triangles 
anchored at a given node by \clm{ti}.
Placing the links not involved in making triangles gives at most a factor
$n$ per link, and therefore, using also Eqs.\equ{A} and \equ{B}, we
get an upper bound
$$
n^{nk} n^{-\ltr/3} 2^{(\alpha +1)n } 16^{\alpha n} 2^{(k+1)n} (2k)^{n 4k^2}~.
$$
The sum over the possible values of $\ltr$ is bounded by $3\alpha n$
times the largest contribution (which occurs for $\ltr=\alpha n/(2k)$),
and we get a bound
$$
3\alpha n\cdot C^n n^{nk} n^{-\alpha n/(6k)} 2^{(\alpha +1)n }~,
\EQ{unstarred}
$$
with 
$$
C\,=\, 16^\alpha 2^{(k+1)} (2k)^{ 4k^2}~.
$$

What about the starred cases? As is visible from Table I, there is
{\em no gain} in the cases 4 and 8.
Since we count everything in terms of links, the case 8 is harmless:
We have no gain, but we also place no link. Thus the bad case is 4.
We will show that case 4 cannot occur too often, and thus a fixed
minimal proportion of
the cases {\em will} give a gain.

In the case $4$ (for a frustrated triangle) we are in the process of
drawing
a triangle in which only an $a$ link is missing. In this case, we
observe the ``history'' of the $c$ link.
Note that the $c$ link {\em originates} at $i$, but its other end has
an index
$m$ which can be greater or less than $i$.
We distinguish several cases:
\setitemindent{F1)}
\item{F1)}$m>i$: Then there are 2 subcases.
\itemitem{F1a)}The link $i\to m$ was placed when a triangle
anchored at some node $i'<i$ was formed. Then it must have been placed
as a $b$ link.
\itemitem{F1b)}The link $i\to m$ was placed when another triangle
anchored at $i$ was formed. Since we begin with the frustrated triangles,
this must have been a frustrated triangle. Note, however, that when the
{\em first} frustrated triangle at $i$ is being placed, this case
cannot occur
and we must begin with the case F1a (or with a case other than case 4).
\item{F2)} $m<i$: There are 4 subcases:
\itemitem{F2a)}The link $i\to m$ was a $b$ link when it was placed.
\itemitem{F2b)}The link was placed as a $c$ link of an $R$ triangle
anchored at $m$.
\itemitem{F2c)}The link was placed as a $c$ link of another $F$ triangle
anchored at $i$.
\itemitem{F2d)}The link was placed as an $a$ link of another frustrated
triangle anchored at $i$. Note, however, that when the
{\em first} frustrated triangle at $i$ is being placed, this case
cannot occur
and we must begin with one of the cases F2a--F2c (or with a case other
than case 4).

In the case $4$ (for a round triangle) we
complete a round triangle with a missing $a$ link. In this case, we
observe the ``history''  of the $b$ link. There is
only one
possibility:
\item{R)}The $b$
link of such a triangle will connect nodes $j\to m$ with both $j$ and
$m$ greater than $i$. Therefore, it can only have been placed as a
$b$ link when we constructed a triangle anchored at a node with label
$i'\le i$.
\medskip

To keep track of the conditions mentioned above, we introduce counters
which ``distribute'' the gain which comes from placing a $b$ or $c$
link
onto those further uses of this link in case 4, where no gain is
possible. To do the bookkeeping, we introduce for every link $ij$ a
counter $c_{ij}$. Each of these is 0 as we start the inductive
procedure to be described below:
$$
c_{ij,0}\,=\,0~,
$$
for all $i\ne j \in \{1,\dots,n\}$.
If a link $ij$ is placed for the first time {\em and} it is a $b$ link
or a $c$ link as it is placed, we
set 
$c_{ij}=\kb$.
Each time a link $ij$ is used as a $c$ link in one of the cases F1a or
F2a--F2c ({\em and only in those})
the counter $c_{ij}$ is reduced by one.
The maximal number of uses in Case 4 of a link placed originally as a
$b$ link is $2k-2$ (used 
$k-1$ times for the $c$ link of an $F$ triangle and another $k-1$
times for the $b$ link of an $R$ triangle). Similarly, a link placed
as a $c$ link can
be used in Case 4 another $k-2$ times as a $c$ link in an $F$ triangle.
Since the number of uses is
less than $\kb$, {\em none of the counters $c_{ij}$ will ever become negative}.

Our last counters keep track of the occurrence of the number of times
we are in case F1b or F2d at a given node $i$. At the beginning of the
induction, we set $a_{i,0}=0$ for all
$i\in \{1,\dots,n\}$. Each time we encounter a case among 1--3,
5--7, F1a, F2a--F2c, or R, at node $i$, we increase $a_i$ by $\ka$. Each
time, we encounter case F1b or F2d, we decrease $a_i$ by 1.
Note that since these latter cases cannot occur more than $k$ times,
and they can not occur for the {\em first} triangle at node $i$, we
conclude that {\em none of the counters $a_i$ ever becomes negative}.

We now prove recursively that at any given
step of the construction after adding triangle $t$,
we have a  bound on the total number of possibilities
which is of the form
$$
N_t\,\le\, n^{q_t(1-\beta)}\; n^{-\sigma
\sum_{rs}{c_{rs,t}/\kcp}}n^{-\rho \sum_{s} a_s/k}\prod _{s=1}^t
\bigl(16 (v_{i_s}+k)^2\bigr) ~,
\EQ{recurse}
$$
where $q_t$ is the number of links already drawn, $i_s$ is the
number of the node at which the $s^{\rm th}$ triangle is anchored  
and $\sigma>0$,
$\beta>0$, and $\rho>0$ will be given later on.
{\em If we can show that there is a positive $\beta $ for which these
inequalities hold, then we have shown a bound of the type of
\clm{upper}}, since none of the counters ever becomes negative.

The recursive proof starts when there is no link and all counters are
equal to 0, hence
the bound is trivially true, ($N_0=1$).

We now explain the action at node $i$. During the construction of the
triangles anchored at $i$, some counters will be
updated, and the bound on the combinatorial factor will evolve
correspondingly.
We now inspect the evolution of the bound during the different possible
actions taken at step $i$. Assume that $t-1$ triangles have been
placed, and that we are placing now triangle $t$ which is anchored at $i_t=i$.

\setitemindent{Case 11}
\item{\bf Case 1:}According to column $R$ or $F$ of Table I,
we have
$$
N_t\,\le\,N_{t-1} n^2~.
$$
But according to the second and third columns of Table I,
the number of links increases by 3, thus $q_t=q_{t-1}+3$.
The link $jm$ was empty at time $t-1$ and will be filled at time
$t$. Therefore, $c_{jm,t-1}=0$ and $c_{jm,t}=\kb$,
and similarly  $c_{[m,i],t-1}=0$ and $c_{[m,i],t}=2k-2$.
Finally, $a_{i,t}=a_{i,t-1}+k$. The other counters are unchanged.
Therefore, we find
$$
\eqalign{
n^2\cdot &\,n^{-\sigma c_{jm,t-1}/\kbp}\,n^{-\sigma c_{[m,i],t-1}/\kbp}
n^{-\rho a_{i,t-1}/\kap}\cr
\,&\le\,n^{3(1-\beta)} \,n^{-\sigma c_{jm,t}/\kbp}\,n^{-\sigma c_{[m,i],t}/\kbp}
n^{-\rho a_{i,t}/\kap}\cdot\,n^{2-3(1-\beta)+2\sigma+\rho} ~,\cr}
\EQ{234}
$$
where $[m,i]=mi$  or $im$ according to the orientation of the link ($R$ or
$F$ case).\myfoot{Note that 
$$
-\sigma c_{jm,t-1}/\kbp=-\sigma
c_{jm,t}/\kbp+\sigma~,
$$ and also 
$$
-\sigma
c_{[m,i],t-1}/\kbp=-\sigma
c_{[m,i],t}/\kbp+\sigma~,
$$
and this exactly compensates the factor $n^{2\sigma}$. Similarly,
$$
-\rho a_{i,t-1}/\ka=-\rho a_{i,t}/\ka +\rho~,
$$ which compensates the
factor
$n^\rho$ at the end of \equ{234}.
}
Since $q_t=q_{t-1}+3$, we see that $N_t$ satisfies the inductive bound
provided the last factor in \equ{234} is $\le 1$, which is the case if
$$
1\,\ge\,3\beta +2\sigma+\rho~.
\EQ{cond01}
$$
\item{\bf Case 2:}Two new links (an $a$ and a $b$) appear and the
combinatorial factor is
$\max(nv_i,nk)\le n(v_i+k)$. The counter $c_{jm}$ is increased by
$\kb$
and the counter $a_i$ is increased by $k$. Also, $q_t=q_{t-1}+2$.
The bound analogous to
\equ{234} is therefore
$$
\eqalign{
n(v_i+k)&\cdot\,n^{-\sigma c_{jm,t-1}/\kbp}\,
n^{-\rho a_{i,t-1}/\kap}\cr
\,&\le\,n^{2(1-\beta)} \,n^{-\sigma c_{jm,t}/\kbp}\,
n^{-\rho a_{i,t}/\kap}\,n^{1-2(1-\beta)+\sigma+\rho}(v_i+k) ~,\cr}
$$
which proves the inductive assumption if 
$$
1\,\ge\,2\beta +\sigma+\rho~.
\EQ{cond02}
$$
\item{\bf Case 3:}The counter $c_{[m,i]}$ is increased by $\kb$ and
$a_i$ is increased by $k$, and the inductive bound is
$$
\eqalign{
nk&\,n^{-\sigma c_{[m,i],t-1}/\kbp}\,
n^{-\rho a_{i,t-1}/\kap}\cr
\,&\le\,n^{2(1-\beta)} \,n^{-\sigma c_{[m,i],t}/\kbp}\,
n^{-\rho a_{i,t}/\kap}\,n^{1-2(1-\beta)+\sigma+\rho}(v_i+k) ~,\cr}
$$
which proves the inductive assumption if \equ{cond02} holds.
\item{\bf Case 5:}The counters $c_{jm}$ and $c_{[m,i]}$ are increased
by $\kb$, and $a_i$ is increased by $k$. Therefore,
$$
\eqalign{
nk\cdot&\,n^{-\sigma c_{jm,t-1}/\kbp}\,n^{-\sigma c_{[m,i],t-1}/\kbp}\,
n^{-\rho a_{i,t-1}/\kap}\cr
\,\le &\,n^{2(1-\beta)}\, n^{-\sigma c_{jm,t}/\kbp}\,n^{-\sigma c_{[m,i],t}/\kbp}\,
n^{-\rho a_{i,t}/\kap}\,n^{1-2(1-\beta)+2\sigma+\rho}(v_i+k) ~,\cr}
$$
which proves the inductive assumption if 
$$
1\,\ge\,2\beta +2\sigma+\rho~.
\EQ{cond03}
$$
\item{\bf Case 6:}The counter $c_{jm}$ is increased by $\kb$, and
$a_i$ is increased by $\ka$. Therefore we get
$$
\eqalign{
n^{0}(v_i+k)^2\cdot&\,n^{-\sigma c_{[jm],t-1}/\kbp}\,
n^{-\rho a_{i,t-1}/\kap}\cr
\,&\le\,n^{1-\beta}\,n^{-\sigma c_{[jm],t}/\kbp}\,
n^{-\rho a_{i,t}/\kap}\,n^{-(1-\beta)+\sigma+\rho}(v_i+k)^2 ~,\cr}
$$
which proves the inductive assumption if 
$$
1\,\ge\,\beta +\sigma+\rho ~.
\EQ{cond04}
$$
\item{\bf Case 7:}The counter $c_{[m,i]}$ is increased by $\kb$, and
$a_i$ is increased by $\ka$. Therefore we get
$$
\eqalign{
n^{0}k^2\cdot&\,n^{-\sigma c_{[m,i],t-1}/\kbp}\,
n^{-\rho a_{i,t-1}/\kap}\cr
\,&\le\,n^{1-\beta}\,n^{-\sigma c_{[m,i],t}/\kbp}\,
n^{-\rho a_{i,t}/\kap}\,n^{-(1-\beta)+\sigma+\rho}(v_i+k)^2 ~,\cr}
$$
which proves the inductive assumption if \equ{cond04} holds.
\item{\bf Case 8:}This case occurs if we want to draw a triangle
anchored at $i$ which has appeared in an earlier phase of the
construction (for example, if its sides are all sides of type $b$ from
triangles anchored at $i'<i$). In this case, no new link, but 1 new
triangle and 
a factor appear
$$
k^2\,\le\,(v_i+k)^2~,
$$
and the inductive assumption evidently holds, since only the number of
triangles increases, but not the number of links.

The conditions we require so far on $\beta $, $\sigma$, and $\rho $ are all
satisfied if
$$
1\,\ge\,3\beta +2\sigma+\rho ~.
\EQ{cond1}
$$

We now come to\hfill\break\noindent
{\bf Case 4:} 
Whenever one of the subcases F1a, F1b, or F2a--F2c applies,
the link $i\to m$ was placed earlier as a $b$ or a $c$ link, and we decrease
the corresponding counter $c_{im}$ by one
unit, and in the case R the counter $c_{mi}$ is decreased. The counter
$a_i$ is unchanged in these cases.
Therefore, we find
that 
$$
n\cdot n ^{-\sigma c_{[m,i],t-1}/\kcp}\,\le\, n^{1-\beta }n
^{-\sigma c_{[m,i],t}/\kcp}\; n^{\beta -\sigma/\kbp}~,
$$
and this proves the inductive assumption provided
$$
\beta \,\le\,\sigma/(2k-2) ~.
\EQ{cond3}
$$
The remaining cases are F1b and F2d. 
In these cases the counter $a_i$ becomes useful: It is decreased by 1
and all other counters are unchanged.
Therefore, we get
$$
n\cdot n ^{-\rho a_{i,t-1}/\ka}\,\le\, n^{1-\beta }n
^{-\rho a_{i,t}/\kap}\; n^{\beta -\rho/\kap}~,
$$
and the inductive assumption holds provided
$$
\beta \,\le\, \rho /\kap~.
\EQ{conda}
$$

We have now discussed all cases. It remains to see that the constants
$\beta $, $\sigma$, and $\rho $ can be chosen consistently.
They have to satisfy \equ{cond1}, \equ{cond3}, and \equ{conda}.
We find that the optimal solution is
$$
\rho = k\beta ~, \quad
\sigma = \kbp \beta ~,\quad \text{and}~~~
1\,=\,3\beta +2\kbp \beta +k\beta ~.
$$   
Thus, we find 
$$
\beta \,=\,{1\over \factor}~,
$$
and the proof of the inductive assumption is complete.

Recall that to draw $t$ triangles we need at least $t/(2k)$ links, and
no more than $3t$ links.
We therefore conclude (combining the bound on $N_t$ with \equ{A} and
\equ{B})  that the number of graphs with $t$
triangles is bounded by
$$
3t\cdot 16^t 2^{t+n} n^{nk-\beta t/(2k)} 2^{(k+1)n} (2k)^{n (2k^2-k)}~,
$$
where the last factor follows from \equ{recurse} and the observation that
the number of triangles anchored at a node is bounded by $2k^2-k$ by \clm{ti}.
The proof of \clm{upper} is complete.\QED

\SECTION{A lower bound for $k$-general graphs}Here, we consider the class
$\GG_{n,k}$ of graphs with $n$ nodes and $kn$ links which can be
placed anywhere we please. (This model is close to the well known
model $G_{n,p}$  where any of the links is chosen with
probability $p=2k/n$.)

\CLAIM{Lemma}{counter}{Fix any $\alpha >0$. The number of graphs with
$\alpha n$ triangles
in the class of graphs with $n$ nodes and $kn$ links is (for large
enough $n$) at least
$$
|\GG_{n,k,\alpha }|\,\ge\,e^{-\OO(n^{2/3})} n^{-(\alpha n)^{2/3}}\;
|\GG_{n,k}|~.
$$
}

\REMARK{5}{This is clearly much larger than the bound of \clm{upper}, and
almost as large as $n^{nk}$, in fact
$$
\lim_{n\to\infty } {\log |\GG_{n,k,\alpha }|\over n\log n}\,=\,k~.
$$
This also proves \equ{11}, since 
$$
|\GG_{n,k}|\,=\,{{n\choose 2} \choose nk}\,\approx\,n^{nk}~.
$$
}

\PROOF Among the $n$ nodes choose $\alpha^{1/3}n^{1/3}$. This can be
done in
$$
{ n \choose \alpha ^{1/3} n^{1/3}}\,\ge\,1
\EQ{l1}
$$
ways.
(Here and throughout the proof we do not worry about the integer parts.)
With these nodes we build a complete graph,
this consumes about $\alpha^{2/3}n^{2/3}$ links and gives
$\alpha n$ triangles (and it can be done in 1 way).
Now among the remaining $n-\alpha^{1/3}n^{1/3}$ nodes
distribute the $kn -\alpha^{2/3}n^{2/3}$ links so that there is no
triangle. The number of ways this 
 can be done can be estimated from below 
 using the well known result on the number 
of graphs without triangles (see [B]), namely this number is bounded below by
$$
\OO(1){{n-\alpha^{1/3}n^{1/3}\choose 2} \choose kn -\alpha^{2/3}n^{2/3}}
\,\ge\, e^{-\OO(n^{2/3})} n^{-(\alpha n)^{2/3}}{{n \choose 2}  \choose nk}~.\EQ{l2}
$$

Combining \equ{l1} and \equ{l2} we get the  lower bound.
\QED

\REMARK{7}{In some sense, this result can be viewed as a complement
to the large deviation results
of Vu[VV], see also [JLR].  
Consider the polynomial associated with triangles in a graph with $n$ nodes:
$$
Y\,=\,\sum_{1\le i<j < m \le n} t_{ij} t_{jm} t_{mi}~.
$$
For this polynomial (in the case of the model $\GG'_{n,p}$ where links
appear with probability $p=k/n$) 
one has
$$
{\bf
E}(Y)\,=\,\OO(1) ~,\quad {\bf E}'(Y)=\OO(n^{-1})~,
$$
with $\bf E$ the expectation in the random set.
Taking Theorem 1.1 in [VV]
and choosing $\lambda =\OO(\alpha ^{1/3} n^{1/2})$ one gets an upper bound of
the form
$$
{\bf P}(|Y|\ge \alpha n)\,\le\, e^{-\OO(\alpha ^{1/3} n^{1/2})}~.
$$
Note that this is consistent with the lower bound of \clm{counter}.}

\SECTION{A lower bound for $k$-out graphs (and for $k$-regular
graphs)}Consider the graphs $\GG_{n,\kout}$ which are of type
$k$-out. Recall that 
$$
|\GG_{n,\kout}|\,\approx\, n^{n k}~.
$$
We will consider graphs $\KK_{k+1}$ which are complete in the sense that
there are ${k+1}$ nodes, and from  each node $k$ out-links are
leaving (to another node of $\KK_{k+1}$). Counting in this case gives
$k+1$ nodes, $k(k+1) $
links (one for each direction) and $8 {k+1\choose 3}$ triangles (the
factor 8 accounting for the 8 ways to use the 6 links on each
triangle, one back and one forth for each pair of nodes).

We now distribute the $\alpha n$ triangles into\myfoot{To simplify
the discussion, which is in terms of orders of magnitude, we assume
that all quotients are integers.}
$$
C_n \,=\,{\alpha n\over 8{k+1 \choose 3}}
$$
disjoint complete graphs $\KK_{k+1}$, and this leaves
$$
R_n\,=\,n-(k+1) {\alpha n\over 8{k+1 \choose 3}}
$$
nodes which will not have been used when making the complete graphs of
type $\KK_{k+1}$.
All links originating from the nodes of the clusters are used up in
forming the $\KK_{k+1}$.
The number of ways to place the $C_n$ complete graphs is
$$
{{n}\choose {{\alpha n(k+1)\over 8{k+1 \choose 3}}}}
{\left ({\alpha n(k+1)\over
8{k+1\choose 3}}\right )!\over (k+1)!^{{\alpha n/
(8{k+1\choose 3})}}\left (\alpha n\over 8{k+1\choose 3}\right )!}
\,\approx\,
n^{ Q_n}~,
\EQ{place}
$$
with
$$
Q_n \,=\,
{\alpha n(k+1)\over 8{k+1\choose 3}}
-{\alpha n\over 8{k+1\choose 3}}~.
$$
The first factor in \equ{place} counts the number of ways to choose
the nodes involved, and the quotient counts the number of ways the $(k+1)C_n$
nodes are grouped into clusters of $k+1$ each.
Since the graphs use $k$ links per node, the graph we can
construct with the remaining $R_n$ nodes will be disjoint from
the $C_n$ clusters, and we want to bound the number of ways in
connecting  the remaining nodes {\em without} adding any triangles.
A lower bound on the number of such graphs is obtained by constructing
again a $k$-out bi-partite graph on the remaining $R_n$ nodes.

The number of ways to place the remaining links is therefore bounded below by
$$
{{R_n/2}\choose k}^{R_n}\,\approx\, n^{S_n}~,
$$
with
$$
S_n\,=\,k\left (
n-{\alpha n(k+1)\over 8{k+1\choose 3}}
\right )~.
$$
Note that we do not insist on making a connected graph.
So we find a lower bound of $E^n n^{T_n}$, where $E$ depends only on
$k$ and
$$
\eqalign{
T_n\,&=\,Q_n+S_n\,=\,n\left (k-{6\alpha \over 8(k+1)k(k-1)}\bigl(k(k+1)-(k+1)+1
\bigr)
\right )\cr
\,&=\,n\left ( k -{3\alpha \over 4k}(1 +{1\over k^2-1})\right )\,=\, nk\left(1-{3\alpha\over 4(k^2-1)}\right)~.\cr
}
\EQ{lll}
$$
\REMARK{99}{The above calculation proves the lower bound for \equ{2}. The
lower bound for \equ{3} is an easy variant, observing the fact that
instead of 8 triangles in a complete graph on 3 nodes in the $k$-out
model there is only 1 in the $k$-regular model.}

\LIKEREMARK{Acknowledgments}We thank E. Moses four helpful
discussions and encouragements. This work was partially supported by
the Fonds National Suisse.

\SECTIONNONR{References}
\widestlabel{[X2]}
{\eightpoint\frenchspacing

\ref
\no{AB}
\by{R. Albert and A.-L. Barab\'asi}
\paper{Statistical mechanics of complex networks}
\jour{Rev. Mod. Phys.}
\pages{47--97}
\yr{2002}
\vol{47}
\endref
\ref
\no{B}
\by{B. Bollobas}
\book{Random Graphs}
\publisher{New York, Academic Press}
\yr{1985}
\endref
\ref
\no{EM}
\by{J.-P. Eckmann and E. Moses}
\paper{Curvature of Co-Links
Uncovers Hidden Thematic Layers in the World Wide Web}
\jour{PNAS}
\inprint
\endref
\ref
\no{JLR}
\by{J. Janson, T. \L uczak, and A. Ruci\'nski}
\paper{An exponential bound for the
probability of nonexistence of a specified subgraph in a random graph}
\bybook{M. Karo\'nski {\it et al.} eds.}
\inbook{Random Graphs 87}
\publisher{Wiley, New York,}
\yr{1990}
\pages{73--87}
\endref
\ref
\no{VV}
\by{H. Vu}
\paper{On the concentration of multi-variate polynomials with
small expectation}
\jour{Random Structures and Algorithms} 
\vol{16}
\pages{344--363}
\yr{2000}
\endref
\bye

%% file: newhead3.tex
%%%%%%%% VERSION OF NEWHEAD WHICH IS NOT BACKWARD COMPATIBLE WITH
%%%%%%%% WITH NEWHEAD2, BUT WHICH WILL BE CLOSER TO LATEX STYLE
%%%%%%%% all macros take Latex style arguments, i.e,
%%%%%%%% macro{arg1}{arg2}  etc.
%%%%%%%% otherwise like newhead2
%%%%%%%%%%%%%%%%%%%%%%%%%%%%%%%%%%%%%%%%%%%%%%%%%%%%%%%%%%%%%%%%%%%%%
\ifx\newheadisloaded\relax\immediate\write16{***already loaded} \else\let\newheadisloaded=\relax\fi
\gdef\isonnarrowscreen{F}
\gdef\PSfonts{T}
%%%%%%%%%%%%%%%%%%%%%%%%%
%%%%layout %%%%%%%%%%%%%%
%%%%%%%%%%%%%%%%%%%%%%%%%
\magnification\magstep1
%for dvips with figures maybe uncomment next line

%%%%%%%%%%%%%%%%%%%%%%%
%%%% new parameters %%%
%%%%%%%%%%%%%%%%%%%%%%%
\newdimen\papwidth
\newdimen\papheight
\newskip\beforesectionskipamount  %how much to skip before section title
\newskip\sectionskipamount %how much to skip after section title
\def\sectionskip{\vskip\sectionskipamount}
\def\beforesectionskip{\vskip\beforesectionskipamount}
%%%%%%%%%%%%%%%%%%%%%%%
%%%% paper %%%%%%%%%%%%
%%%%%%%%%%%%%%%%%%%%%%%
\papwidth=16truecm
\if F\isonnarrowscreen
\papheight=22truecm
\voffset=0.4truecm
\hoffset=0.4truecm
\else
\papheight=22truecm
\voffset=-1.5truecm
\hoffset=-1truecm
\fi
%%%%%%%install variables%%%%%%%%%%%%%%%%%%%
\hsize=\papwidth
\vsize=\papheight
%%%%%%%%%%%%%%%%%%%%%%%%%
%%%%%% headline %%%%%%%%%
%%%%%%%%%%%%%%%%%%%%%%%%%
\catcode`\@=11
\ifx\amstexloaded@\relax
\else
\nopagenumbers
\headline={\ifnum\pageno>1 {\hss\tenrm-\ \folio\ -\hss} \else
{\hfill}\fi}
\fi
\catcode`\@=\active
%%%%%%%%%%%%%%%%%%%%%%%%%
\newdimen\texpscorrection
\texpscorrection=0.15truecm %must be 0.15truecm in ps_fonts
%%%%%%%%%%%%%%%%%%%%%%%%
%%%%%%% fontsizes %%%%%%
%%%%%%%%%%%%%%%%%%%%%%%%

\def\sectionsize{\twelvepoint}
\def\sectiontype{\bf}
\def\subsectionsize{}
\def\subsectiontype{\bf}
\def\em{\sl}%will be italic in reality
%%%%%%%%%%%%%%%%%%%%%
\newfam\truecmsy
\newfam\truecmr
\newfam\msbfam
\newfam\scriptfam
\newfam\frakfam
\newfam\frakbfam

%%%%%%%%%%%%%%%%%%%%%%%%%%%%%%%%%%%%%
\newskip\ttglue 
%%%%%%%%%%%%%%%%%%%%%%%%%%%%%%%%%%%%%%%%%%%
% Font for LINUX
%%%%%%%%%%%%%%%%%%%%%%%%%%%%%%%%%%%%%%%%%%%%%
\if T\isonnarrowscreen
\papheight=11.5truecm
\fi
\if F\PSfonts
% Times-Roman
\font\twelverm=cmr12
\font\tenrm=cmr10
%\font\ninerm=cmr9
\font\eightrm=cmr8
\font\sevenrm=cmr7
\font\sixrm=cmr6
\font\fiverm=cmr5

% Times-Bold
\font\twelvebf=cmbx12
\font\tenbf=cmbx10
%\font\ninebf=cmbx9
\font\eightbf=cmbx8
\font\sevenbf=cmbx7
\font\sixbf=cmbx6
\font\fivebf=cmbx5

% Times-Italic
\font\twelveit=cmti12
\font\tenit=cmti10
%\font\nineit=cmti9
\font\eightit=cmti8
\font\sevenit=cmti7
\font\sixit=cmti6
\font\fiveit=cmti5

% Times-Oblique(slanted)
\font\twelvesl=cmsl12
\font\tensl=cmsl10
%\font\ninesl=cmsl9
\font\eightsl=cmsl8
\font\sevensl=cmsl7
\font\sixsl=cmsl6
\font\fivesl=cmsl5

% Math-Italic
\font\twelvei=cmmi12
\font\teni=cmmi10
%\font\ninei=cmmi9
\font\eighti=cmmi8
\font\seveni=cmmi7
\font\sixi=cmmi6
\font\fivei=cmmi5

% Math-Symbols
\font\twelvesy=cmsy10	at	12pt
\font\tensy=cmsy10
%\font\ninesy=cmsy9
\font\eightsy=cmsy8
\font\sevensy=cmsy7
\font\sixsy=cmsy6
\font\fivesy=cmsy5
\font\twelvetruecmsy=cmsy10	at	12pt
\font\tentruecmsy=cmsy10
%\font\ninetruecmsy=cmsy9
\font\eighttruecmsy=cmsy8
\font\seventruecmsy=cmsy7
\font\sixtruecmsy=cmsy6
\font\fivetruecmsy=cmsy5

% CM-Roman
\font\twelvetruecmr=cmr12
\font\tentruecmr=cmr10
%\font\ninetruecmr=cmr9
\font\eighttruecmr=cmr8
\font\seventruecmr=cmr7
\font\sixtruecmr=cmr6
\font\fivetruecmr=cmr5

% Math-Boldfaces
\font\twelvebf=cmbx12
\font\tenbf=cmbx10
%\font\ninebf=cmbx9
\font\eightbf=cmbx8
\font\sevenbf=cmbx7
\font\sixbf=cmbx6
\font\fivebf=cmbx5

% Teletype
\font\twelvett=cmtt12
\font\tentt=cmtt10
%\font\ninett=cmtt9
\font\eighttt=cmtt8

% Big Math Symbols
\font\twelveex=cmex10	at	12pt
\font\tenex=cmex10
%\font\nineex=cmex9

% AMS Math Symbols
\font\twelvemsb=msbm10	at	12pt
\font\tenmsb=msbm10
%\font\ninemsb=msbm9
\font\eightmsb=msbm8
\font\sevenmsb=msbm7
\font\sixmsb=msbm6
\font\fivemsb=msbm5

%%% Fraktur
%\newfam\frakfam

\font\tenfrm=eufm10
%\font\ninefrm=eufm9
\font\eightfrm=eufm8
\font\sevenfrm=eufm7
\font\sixfrm=eufm6
\font\fivefrm=eufm5
%%
%%% Bold Fraktur
%%\newfam\frakbfam

\font\tenfrb=eufb10
%\font\ninefrb=eufb9
\font\eightfrb=eufb8
\font\sevenfrb=eufb7
\font\sixfrb=eufb6
\font\fivefrb=eufb5
%%
% Script-Faces
\font\twelvescr=eusm10 at 12pt
\font\tenscr=eusm10
%\font\ninescr=eusm9
\font\eightscr=eusm8
\font\sevenscr=eusm7
\font\sixscr=eusm6
\font\fivescr=eusm5
\fi
\if T\PSfonts
%%%%%%%%%%%%%%%%%%%%%%%%%%%%%%%%%%%%%%
% Font mapping for postscript fonts.
%%%%%%%%%%%%%%%%%%%%%%%%%%%%%%%%%%%%%%
% Times-Roman
\font\twelverm=ptmr	at	12pt
\font\tenrm=ptmr	at	10pt
%\font\ninerm=ptmr	at	9pt
\font\eightrm=ptmr	at	8pt
\font\sevenrm=ptmr	at	7pt
\font\sixrm=ptmr	at	6pt
\font\fiverm=ptmr	at	5pt

% Times-Bold
\font\twelvebf=ptmb	at	12pt
\font\tenbf=ptmb	at	10pt
%\font\ninebf=ptmb	at	9pt
\font\eightbf=ptmb	at	8pt
\font\sevenbf=ptmb	at	7pt
\font\sixbf=ptmb	at	6pt
\font\fivebf=ptmb	at	5pt

% Times-Italic
\font\twelveit=ptmri	at	12pt
\font\tenit=ptmri	at	10pt
%\font\nineit=ptmri	at	9pt
\font\eightit=ptmri	at	8pt
\font\sevenit=ptmri	at	7pt
\font\sixit=ptmri	at	6pt
\font\fiveit=ptmri	at	5pt

% Times-Oblique(slanted)
\font\twelvesl=ptmro	at	12pt
\font\tensl=ptmro	at	10pt
%\font\ninesl=ptmro	at	9pt
\font\eightsl=ptmro	at	8pt
\font\sevensl=ptmro	at	7pt
\font\sixsl=ptmro	at	6pt
\font\fivesl=ptmro	at	5pt

% Math-Italic
\font\twelvei=cmmi12
\font\teni=cmmi10
%\font\ninei=cmmi9
\font\eighti=cmmi8
\font\seveni=cmmi7
\font\sixi=cmmi6
\font\fivei=cmmi5

% Math-Symbols
\font\twelvesy=cmsy10	at	12pt
\font\tensy=cmsy10
%\font\ninesy=cmsy9
\font\eightsy=cmsy8
\font\sevensy=cmsy7
\font\sixsy=cmsy6
\font\fivesy=cmsy5
\font\twelvetruecmsy=cmsy10	at	12pt
\font\tentruecmsy=cmsy10
%\font\ninetruecmsy=cmsy9
\font\eighttruecmsy=cmsy8
\font\seventruecmsy=cmsy7
\font\sixtruecmsy=cmsy6
\font\fivetruecmsy=cmsy5

% CM-Roman
\font\twelvetruecmr=cmr12
\font\tentruecmr=cmr10
%\font\ninetruecmr=cmr9
\font\eighttruecmr=cmr8
\font\seventruecmr=cmr7
\font\sixtruecmr=cmr6
\font\fivetruecmr=cmr5

% Math-Boldfaces
%\font\twelvebf=cmbx12
%\font\tenbf=cmbx10
%\font\ninebf=cmbx9
%\font\eightbf=cmbx8
%\font\sevenbf=cmbx7
%\font\sixbf=cmbx6
%\font\fivebf=cmbx5

% Teletype
\font\twelvett=cmtt12
\font\tentt=cmtt10
%\font\ninett=cmtt9
\font\eighttt=cmtt8

% Big Math Symbols
\font\twelveex=cmex10	at	12pt
\font\tenex=cmex10
%\font\nineex=cmex9

% AMS Math Symbols
\font\twelvemsb=msbm10	at	12pt
\font\tenmsb=msbm10
%\font\ninemsb=msbm9
\font\eightmsb=msbm8
\font\sevenmsb=msbm7
\font\sixmsb=msbm6
\font\fivemsb=msbm5

%%% Fraktur
%%\newfam\frakfam

\font\tenfrm=eufm10
%\font\ninefrm=eufm9
\font\eightfrm=eufm8
\font\sevenfrm=eufm7
\font\sixfrm=eufm6
\font\fivefrm=eufm5
%%
%%% Bold Fraktur
%%\newfam\frakbfam

\font\tenfrb=eufb10
%\font\ninefrb=eufb9
\font\eightfrb=eufb8
\font\sevenfrb=eufb7
\font\sixfrb=eufb6
\font\fivefrb=eufb5
%%
% Script-Faces
\font\twelvescr=eusm10 at 12pt
\font\tenscr=eusm10
%\font\ninescr=eusm9
\font\eightscr=eusm8
\font\sevenscr=eusm7
\font\sixscr=eusm6
\font\fivescr=eusm5
\fi
%%%%%%%%%%%%%%%%%%%%%%%%%
%%%%preloaded fonts%%%%%%
%%%%%%%%%%%%%%%%%%%%%%%%%
\def\eightpoint{\def\rm{\fam0\eightrm}%
\textfont0=\eightrm
  \scriptfont0=\sixrm
  \scriptscriptfont0=\fiverm 
\textfont1=\eighti
  \scriptfont1=\sixi
  \scriptscriptfont1=\fivei 
\textfont2=\eightsy
  \scriptfont2=\sixsy
  \scriptscriptfont2=\fivesy 
\textfont3=\tenex
  \scriptfont3=\tenex
  \scriptscriptfont3=\tenex 
\textfont\itfam=\eightit
  \scriptfont\itfam=\sixit
  \scriptscriptfont\itfam=\fiveit 
  \def\it{\fam\itfam\eightit}%
\textfont\slfam=\eightsl
  \scriptfont\slfam=\sixsl
  \scriptscriptfont\slfam=\fivesl 
  \def\sl{\fam\slfam\eightsl}%
\textfont\ttfam=\eighttt
  \def\tt{\fam\ttfam\eighttt}%
\textfont\bffam=\eightbf
  \scriptfont\bffam=\sixbf
  \scriptscriptfont\bffam=\fivebf
  \def\bf{\fam\bffam\eightbf}%
\textfont\frakfam=\eightfrm
  \scriptfont\frakfam=\sixfrm
  \scriptscriptfont\frakfam=\fivefrm
  \def\frak{\fam\frakfam\eightfrm}%
\textfont\frakbfam=\eightfrb
  \scriptfont\frakbfam=\sixfrb
  \scriptscriptfont\frakbfam=\fivefrb
  \def\bfrak{\fam\frakbfam\eightfrb}%
\textfont\scriptfam=\eightscr
  \scriptfont\scriptfam=\sixscr
  \scriptscriptfont\scriptfam=\fivescr
  \def\script{\fam\scriptfam\eightscr}%
\textfont\msbfam=\eightmsb
  \scriptfont\msbfam=\sixmsb
  \scriptscriptfont\msbfam=\fivemsb
  \def\bb{\fam\msbfam\eightmsb}%
\textfont\truecmr=\eighttruecmr
  \scriptfont\truecmr=\sixtruecmr
  \scriptscriptfont\truecmr=\fivetruecmr
  \def\truerm{\fam\truecmr\eighttruecmr}%
\textfont\truecmsy=\eighttruecmsy
  \scriptfont\truecmsy=\sixtruecmsy
  \scriptscriptfont\truecmsy=\fivetruecmsy
\tt \ttglue=.5em plus.25em minus.15em 
\normalbaselineskip=9pt
\setbox\strutbox=\hbox{\vrule height7pt depth2pt width0pt}%
\normalbaselines
\rm
}

\def\tenpoint{\def\rm{\fam0\tenrm}%
\textfont0=\tenrm
  \scriptfont0=\sevenrm
  \scriptscriptfont0=\fiverm 
\textfont1=\teni
  \scriptfont1=\seveni
  \scriptscriptfont1=\fivei 
\textfont2=\tensy
  \scriptfont2=\sevensy
  \scriptscriptfont2=\fivesy 
\textfont3=\tenex
  \scriptfont3=\tenex
  \scriptscriptfont3=\tenex 
\textfont\itfam=\tenit
  \scriptfont\itfam=\sevenit
  \scriptscriptfont\itfam=\fiveit 
  \def\it{\fam\itfam\tenit}%
\textfont\slfam=\tensl
  \scriptfont\slfam=\sevensl
  \scriptscriptfont\slfam=\fivesl 
  \def\sl{\fam\slfam\tensl}%
\textfont\ttfam=\tentt
  \def\tt{\fam\ttfam\tentt}%
\textfont\bffam=\tenbf
  \scriptfont\bffam=\sevenbf
  \scriptscriptfont\bffam=\fivebf
  \def\bf{\fam\bffam\tenbf}%
\textfont\frakfam=\tenfrm
  \scriptfont\frakfam=\sevenfrm
  \scriptscriptfont\frakfam=\fivefrm
  \def\frak{\fam\frakfam\tenfrm}%
\textfont\frakbfam=\tenfrb
  \scriptfont\frakbfam=\sevenfrb
  \scriptscriptfont\frakbfam=\fivefrb
  \def\bfrak{\fam\frakbfam\tenfrb}%
\textfont\scriptfam=\tenscr
  \scriptfont\scriptfam=\sevenscr
  \scriptscriptfont\scriptfam=\fivescr
  \def\script{\fam\scriptfam\tenscr}%
\textfont\msbfam=\tenmsb
  \scriptfont\msbfam=\sevenmsb
  \scriptscriptfont\msbfam=\fivemsb
  \def\bb{\fam\msbfam\tenmsb}%
\textfont\truecmr=\tentruecmr
  \scriptfont\truecmr=\seventruecmr
  \scriptscriptfont\truecmr=\fivetruecmr
  \def\truerm{\fam\truecmr\tentruecmr}%
\textfont\truecmsy=\tentruecmsy
  \scriptfont\truecmsy=\seventruecmsy
  \scriptscriptfont\truecmsy=\fivetruecmsy
\tt \ttglue=.5em plus.25em minus.15em 
\normalbaselineskip=12pt
\setbox\strutbox=\hbox{\vrule height8.5pt depth3.5pt width0pt}%
\normalbaselines
\rm
}

\def\twelvepoint{\def\rm{\fam0\twelverm}%
\textfont0=\twelverm
  \scriptfont0=\tenrm
  \scriptscriptfont0=\eightrm 
\textfont1=\twelvei
  \scriptfont1=\teni
  \scriptscriptfont1=\eighti 
\textfont2=\twelvesy
  \scriptfont2=\tensy
  \scriptscriptfont2=\eightsy 
\textfont3=\twelveex
  \scriptfont3=\twelveex
  \scriptscriptfont3=\twelveex 
\textfont\itfam=\twelveit
  \scriptfont\itfam=\tenit
  \scriptscriptfont\itfam=\eightit 
  \def\it{\fam\itfam\twelveit}%
\textfont\slfam=\twelvesl
  \scriptfont\slfam=\tensl
  \scriptscriptfont\slfam=\eightsl 
  \def\sl{\fam\slfam\twelvesl}%
\textfont\ttfam=\twelvett
  \def\tt{\fam\ttfam\twelvett}%
\textfont\bffam=\twelvebf
  \scriptfont\bffam=\tenbf
  \scriptscriptfont\bffam=\eightbf
  \def\bf{\fam\bffam\twelvebf}%
%%\textfont\frakfam=\twelvefrm
%%  \scriptfont\frakfam=\tenfrm
%%  \scriptscriptfont\frakfam=\eightfrm
%%  \def\frak{\fam\frakfam\twelvefrm}%
%%\textfont\frakbfam=\twelvefrb
%%  \scriptfont\frakbfam=\tenfrb
%%  \scriptscriptfont\frakbfam=\eightfrb
%%  \def\bfrak{\fam\frakbfam\twelvefrb}%
\textfont\scriptfam=\twelvescr
  \scriptfont\scriptfam=\tenscr
  \scriptscriptfont\scriptfam=\eightscr
  \def\script{\fam\scriptfam\twelvescr}%
\textfont\msbfam=\twelvemsb
  \scriptfont\msbfam=\tenmsb
  \scriptscriptfont\msbfam=\eightmsb
  \def\bb{\fam\msbfam\twelvemsb}%
\textfont\truecmr=\twelvetruecmr
  \scriptfont\truecmr=\tentruecmr
  \scriptscriptfont\truecmr=\eighttruecmr
  \def\truerm{\fam\truecmr\twelvetruecmr}%
\textfont\truecmsy=\twelvetruecmsy
  \scriptfont\truecmsy=\tentruecmsy
  \scriptscriptfont\truecmsy=\eighttruecmsy
\tt \ttglue=.5em plus.25em minus.15em 
\setbox\strutbox=\hbox{\vrule height7pt depth2pt width0pt}%
\normalbaselineskip=15pt
\normalbaselines
\rm
}
%
%%%%%constant subscript positions%%%%%
\fontdimen16\tensy=2.7pt
%\fontdimen13\tensy=2.7pt
\fontdimen13\tensy=4.3pt
\fontdimen17\tensy=2.7pt
\fontdimen14\tensy=4.3pt
\fontdimen18\tensy=4.3pt
\fontdimen16\eightsy=2.7pt
\fontdimen13\eightsy=4.3pt
\fontdimen17\eightsy=2.7pt
\fontdimen14\eightsy=4.3pt
\fontdimen18\sevensy=4.3pt
\fontdimen16\sevensy=1.8pt
\fontdimen13\sevensy=4.3pt
\fontdimen17\sevensy=2.7pt
\fontdimen14\sevensy=4.3pt
\fontdimen18\sevensy=4.3pt
%
%%%%%%%%%%%%%%%%%%%%%%%%%%%%%%%%%%%%%%%%%%%%%%%%%%%%%%%%%%%%%
%%%%%%%%%%%%%% redefine some math so that it is cmr %%%%%%%%%
%%%%%%%%%%%%%%%%%%%%%%%%%%%%%%%%%%%%%%%%%%%%%%%%%%%%%%%%%%%%%
\def\hexnumber#1{\ifcase#1 0\or1\or2\or3\or4\or5\or6\or7\or8\or9\or
 A\or B\or C\or D\or E\or F\fi}
\mathcode`\=="3\hexnumber\truecmr3D
\mathchardef\not="3\hexnumber\truecmsy36
\mathcode`\+="2\hexnumber\truecmr2B
\mathcode`\(="4\hexnumber\truecmr28
\mathcode`\)="5\hexnumber\truecmr29
\mathcode`\!="5\hexnumber\truecmr21
\mathcode`\(="4\hexnumber\truecmr28
\mathcode`\)="5\hexnumber\truecmr29
%\chardef`,="0\hexnum\truecmr3B

\def\Phi{\mathchar"0\hexnumber\truecmr08 }
\def\Gamma {\mathchar"0\hexnumber\truecmr00 }
\def\Delta {\mathchar"0\hexnumber\truecmr01 }
\def\Theta {\mathchar"0\hexnumber\truecmr02 }
\def\Lambda{\mathchar"0\hexnumber\truecmr03 }
\def\Xi {\mathchar"0\hexnumber\truecmr04 }
\def\Pi{\mathchar"0\hexnumber\truecmr05 }
\def\Sigma{\mathchar"0\hexnumber\truecmr06 }
\def\Upsilon {\mathchar"0\hexnumber\truecmr07 }
\def\Phi {\mathchar"0\hexnumber\truecmr08 }
\def\Psi {\mathchar"0\hexnumber\truecmr09 }
\def\Omega{\mathchar"0\hexnumber\truecmr0A }
%%%%%%%%%%%%%%%%%%%%%%%%%%%%%%%%%%%%%%%%%%%%%%
%%% macros  for cross reference %%%%%%%%%%%%%%
%%%%%%%%%%%%%%%%%%%%%%%%%%%%%%%%%%%%%%%%%%%%%%
%%
%%  counters %%%
%%  
\newcount\EQNcount \EQNcount=1
\newcount\CLAIMcount \CLAIMcount=1
\newcount\SECTIONcount \SECTIONcount=0
\newcount\SUBSECTIONcount \SUBSECTIONcount=1
%%
%% defining the symbolic value
%%
\def\ifff#1#2#3{\ifundefined{#1#2}%
\expandafter\xdef\csname #1#2\endcsname{#3}\else%
\immediate\write16{!!!!!doubly defined #1,#2}\fi}
\def\NEWDEF#1#2#3{\ifff{#1}{#2}{#3}}
\def\actualnumber{\number\SECTIONcount}
\def\EQ#1{\lmargin{#1}\eqno\tageck{#1}}
\def\NR#1{&\lmargin{#1}\tageck{#1}\cr}  %the same as &\tageck{xx}\cr in eqalignno
\def\tageck#1{\lmargin{#1}({\rm \actualnumber}.\number\EQNcount)
 \NEWDEF{e}{#1}{(\actualnumber.\number\EQNcount)}
\global\advance\EQNcount by 1
%\immediate\write16{ EQ \equ{#1}:#1  }
}

%%%% the actual macro %%%%%%
\def\CLAIM#1#2#3\par{
\vskip.1in\medbreak\noindent
{\lmargin{#2}\bf #1\ \actualnumber.\number\CLAIMcount.} {\sl #3}\par
\NEWDEF{c}{#2}{#1\ \actualnumber.\number\CLAIMcount}
\global\advance\CLAIMcount by 1
\ifdim\lastskip<\medskipamount
\removelastskip\penalty55\medskip\fi}
\def\CLAIMNONR #1#2#3\par{
\vskip.1in\medbreak\noindent
{\lmargin{#2}\bf #1.} {\sl #3}\par
\NEWDEF{c}{#2}{#1}
\global\advance\CLAIMcount by 1
\ifdim\lastskip<\medskipamount
\removelastskip\penalty55\medskip\fi}
%%%%%%%%%%%%%%%%%%%%%%%%%%%%%%%%%%%%%%%%%%%%
\def\SECTION#1{\vskip0pt plus.2\vsize\penalty-75
    \vskip0pt plus -.2\vsize
    \global\advance\SECTIONcount by 1
    \beforesectionskip\noindent
{\sectionsize\sectiontype \actualnumber.\ #1}
    \EQNcount=1
    \CLAIMcount=1
    \SUBSECTIONcount=1
    \nobreak\sectionskip\noindent}
\def\SECTIONNONR#1{\vskip0pt plus.3\vsize\penalty-75
    \vskip0pt plus -.3\vsize
    \global\advance\SECTIONcount by 1
    \beforesectionskip\noindent
{\sectionsize\sectiontype  #1}
     \EQNcount=1
     \CLAIMcount=1
     \SUBSECTIONcount=1
     \nobreak\sectionskip\noindent}
\def\SUBSECTION#1{\vskip0pt plus.2\vsize\penalty-75%
    \vskip0pt plus -.2\vsize%
    \beforesectionskip\noindent%
{\subsectionsize\subsectiontype \actualnumber.\number\SUBSECTIONcount.\ #1}
    \global\advance\SUBSECTIONcount by 1
    \nobreak\sectionskip\noindent}
\def\SUBSECTIONNONR#1\par{\vskip0pt plus.2\vsize\penalty-75
    \vskip0pt plus -.2\vsize
\beforesectionskip\noindent
{\subsectionsize\subsectiontype #1}
    \nobreak\sectionskip\noindent\noindent}
%%
%%  referring to something
%%
\def\ifundefined#1{\expandafter\ifx\csname#1\endcsname\relax}
\def\equ#1{\ifundefined{e#1}$\spadesuit$#1\else\csname e#1\endcsname\fi}
\def\clm#1{\ifundefined{c#1}$\spadesuit$#1\else\csname c#1\endcsname\fi}
\def\sec#1{\ifundefined{s#1}$\spadesuit$#1
\else Section \csname s#1\endcsname\fi}
\def\lab#1#2{\ifundefined{#1#2}$\spadesuit$#2\else\csname #1#2\endcsname\fi}
\def\fig#1{\ifundefined{fig#1}$\spadesuit$#1\else\csname fig#1\endcsname\fi}
%%%%%%%%%%%%%TITLE PAGE%%%%%%%%%%%%%%%%%%%%
\let\endarg=\par
\def\finish{\def\endarg{\par\endgroup}}
\def\start{\endarg\begingroup}

 \def\beginFROM{\start\parskip=0pt\vskip\baselineskip
\def\finish{\def\endarg{\egroup\par\endgroup}}
  \vbox\bgroup\obeylines\eightpoint\em\finish}

\def\ABSTRACT#1\par{
\vskip 1in {\noindent\sectionsize\sectiontype Abstract.} #1 \par}

%%%%%%%%%%% The today mechanism %%%%%%%%%%%%%%%%%
\def\TODAY{\number\day~\ifcase\month\or January \or February \or March \or
April \or May \or June
\or July \or August \or September \or October \or November \or December \fi
\number\year\timecount=\number\time
\divide\timecount by 60
}
\newcount\timecount
\def\DRAFT{\def\lmargin##1{\strut\vadjust{\kern-\strutdepth
\vtop to \strutdepth{
\baselineskip\strutdepth\vss\rlap{\kern-1.2 truecm\eightpoint{##1}}}}}%1.2
\font\footfont=cmti7
\footline={{\footfont \hfil File:\jobname, \TODAY,  \number\timecount h}}
}
%%%subitem an item in a vbox%%%%
\newbox\strutboxJPE
\setbox\strutboxJPE=\hbox{\strut}
\def\subitem#1#2\par{\vskip\baselineskip\vskip-\ht\strutboxJPE{\item{#1}#2}}
\gdef\strutdepth{\dp\strutbox}
\def\lmargin#1{}
\def\hexnumber#1{\ifcase#1 0\or1\or2\or3\or4\or5\or6\or7\or8\or9\or
 A\or B\or C\or D\or E\or F\fi}
\textfont\msbfam=\tenmsb
\scriptfont\msbfam=\sevenmsb
\scriptscriptfont\msbfam=\fivemsb
\mathchardef\varkappa="0\hexnumber\msbfam7B%
%%%%%%%%%%%%%%%%%%%%%%%%%%%%%%%%%%%%%%%%%%%%%%%%%%%%%%%%
%%%%%%%%%%  Figures %%%%%%%%%%%%%%%%%%%%%%%%%%%%%%%%%%%%
%%%%%%%%%%%%%%%%%%%%%%%%%%%%%%%%%%%%%%%%%%%%%%%%%%%%%%%%
\newcount\FIGUREcount \FIGUREcount=0
\newdimen\figcenter
\def\definefigure#1{\global\advance\FIGUREcount by 1%
\NEWDEF{fig}{#1}{Fig.\ \number\FIGUREcount}
\immediate\write16{  FIG \number\FIGUREcount : #1}}
%%%%%%%%%%%%%%%%%%%%%%%%%%%%%%%%%%%%%%%%%%%%%%%
%figure 1=psfile=NAME 2=height (in cm) 3=width (in cm) 4=caption  
%%%%%%%%%%%%%%%%%%%%%%%%%%%%%%%%%%%%%%%%%%%%%%%%%%%%%%%
\def\figure#1#2#3#4\cr{\null%
\definefigure{#1}
{\goodbreak\figcenter=\hsize\relax
\advance\figcenter by -#3truecm
\divide\figcenter by 2
\midinsert\vskip #2truecm\noindent\hskip\figcenter
\includegraphics{#1}\vskip 0.8truecm\noindent \vbox{\eightpoint\noindent
{\bf\fig{#1}}: #4}\endinsert}}
%%%%%%%%%%%%%%%%%%%%%%%%%%%%%%%%%%%%%%%%%%%%%%%%%%%%%%%%%
%figurewithtex 1=psfile=NAME 2=texfile 3=height (in cm) 4=width
%(in cm) 5=caption
%%%%%%%%%%%%%%%%%%%%%%%%%%%%%%%%%%%%%%%%%%%%%%%%%%%%%%%%%
\def\figurewithtex#1#2#3#4#5\cr{\null%
\definefigure{#1}
{\goodbreak\figcenter=\hsize\relax
\advance\figcenter by -#4truecm
\divide\figcenter by 2
\midinsert\vskip #3truecm\noindent\hskip\figcenter
\includegraphics{#1}{\hskip\texpscorrection\input #2 }\vskip 0.8truecm\noindent \vbox{\eightpoint\noindent
{\bf\fig{#1}}: #5}\endinsert}}
%%%%%%%%%%%%%%%%%%%%%%%%%%%%%%%%%%%%%%%%%%%%%%%%%%%%%%%%%%%%%%%%%%
%figurewithtexplus 1=psfile=NAME 2=texfile 3=height (in cm) 4=width
%(in cm) 5=dist figure-caption 6=caption
%%%%%%%%%%%%%%%%%%%%%%%%%%%%%%%%%%%%%%%%%%%%%%%%%%%%%%%%%%%%%%%%%%
\def\figurewithtexplus#1#2#3#4#5#6\cr{\null%
\definefigure{#1}
{\goodbreak\figcenter=\hsize\relax
\advance\figcenter by -#4truecm
%\advance\figcenter by -#4truecm
\divide\figcenter by 2
\midinsert\vskip #3truecm\noindent\hskip\figcenter
\includegraphics{#1}{\hskip\texpscorrection\input #2 }\vskip #5truecm\noindent \vbox{\eightpoint\noindent
{\bf\fig{#1}}: #6}\endinsert}}
%%%%%%%%%%%%%%%%%%%%%%%%%%%%%%%%%%%%%%%%%%%%%%%%%%%%%%%
\catcode`@=11
\def\footnote#1{\let\@sf\empty % parameter #2 (the text) is read later
  \ifhmode\edef\@sf{\spacefactor\the\spacefactor}\/\fi
  #1\@sf\vfootnote{#1}}
\def\vfootnote#1{\insert\footins\bgroup\eightpoint
  \interlinepenalty\interfootnotelinepenalty
  \splittopskip\ht\strutbox % top baseline for broken footnotes
  \splitmaxdepth\dp\strutbox \floatingpenalty\@MM
  \leftskip\z@skip \rightskip\z@skip \spaceskip\z@skip \xspaceskip\z@skip
  \textindent{#1}\footstrut\futurelet\next\fo@t}
\def\fo@t{\ifcat\bgroup\noexpand\next \let\next\f@@t
  \else\let\next\f@t\fi \next}
\def\f@@t{\bgroup\aftergroup\@foot\let\next}
\def\f@t#1{#1\@foot}
\def\@foot{\strut\egroup}
\def\footstrut{\vbox to\splittopskip{}}
\skip\footins=\bigskipamount % space added when footnote is present
\count\footins=1000 % footnote magnification factor (1 to 1)
\dimen\footins=8in % maximum footnotes per page
\catcode`@=12 % at signs are no longer letters
%%%%%%%%%%%%%%%%%%%%%%%
%%%  math symbols %%%%%
%%%%%%%%%%%%%%%%%%%%%%%

\def\OO{{\script O}}

%%%%%%%%%%%%%%%%%other%%%%%%%%%%%%%%%%%%%%

\def\QEDD{\hfill\smallskip
         \line{$\hfill{\vcenter{\vbox{\hrule height 0.2pt
	\hbox{\vrule width 0.2pt height 1.3ex \kern 1.3ex
		\vrule width 0.2pt}
	\hrule height 0.2pt}}}$
               \ \ \ \ \ \ }
         \bigskip}
\def\QED{$\hfill{\vcenter{\vbox{\hrule height 0.2pt
	\hbox{\vrule width 0.2pt height 1.3ex \kern 1.3ex
		\vrule width 0.2pt}
	\hrule height 0.2pt}}}$\bigskip}

\def\natural{{\bf N}}

\def\PROOF{\medskip\noindent{\bf Proof.\ }}
\def\REMARK{\medskip\noindent{\bf Remark.\ }}
\def\LIKEREMARK#1{\medskip\noindent{\bf #1.\ }}
%%%%%%%%%%%%%%%%%%%%%%%%%%%%%%%%%%%%%%%%%%%%%%%%%%%
%%%% paragraphs ...                        %%%%%%%%
%%%%%%%%%%%%%%%%%%%%%%%%%%%%%%%%%%%%%%%%%%%%%%%%%%%
\normalbaselineskip=5.25mm
\baselineskip=5.25mm
\parskip=10pt
\beforesectionskipamount=24pt plus8pt minus8pt
\sectionskipamount=3pt plus1pt minus1pt
%\beforesectionskipamount=42pt plus5pt minus2pt
%\sectionskipamount=1truecm
%\overfullrule=0pt
%\hfuzz=2pt
\def\em{\it}
%%%%%%%%%%%%%%%%%%%%%%%%%%%%%%%%%%%%%%%%%%%%%%%%
% choice of default layout %%%%%%%%%%%%%%%%%%%%%
\tenpoint
\null
%%%%%%%%%%%%%%%%%%%%%%%%%%%%%%%%%%%%%%%%%%%%%%%
%%%%%%% exit here if amstex %%%%%%%%%%%%%%%%%%%
\catcode`\@=11
\ifx\amstexloaded@\relax\catcode`\@=\active
 \fi
\catcode`\@=\active
%%%%%%%%%%%%%%%%BIBLIOGRAPHY%%%%%%%%%%%%%%%%%%%%
%%%%%%%%%%%%%%%%%%%%%%%%%%%%%%%%%%%%%%%%%%%%%%%%
\def\period{\unskip.\spacefactor3000 { }}
%
% ...invisible stuff
%
\newbox\noboxJPE
\newbox\byboxJPE
\newbox\paperboxJPE
\newbox\yrboxJPE
\newbox\jourboxJPE
\newbox\pagesboxJPE
\newbox\volboxJPE
\newbox\preprintboxJPE
\newbox\toappearboxJPE
\newbox\bookboxJPE
\newbox\bybookboxJPE
\newbox\publisherboxJPE
\newbox\inprintboxJPE
\def\refclearJPE{
   \setbox\noboxJPE=\null             \gdef\isnoJPE{F}
   \setbox\byboxJPE=\null             \gdef\isbyJPE{F}
   \setbox\paperboxJPE=\null          \gdef\ispaperJPE{F}
   \setbox\yrboxJPE=\null             \gdef\isyrJPE{F}
   \setbox\jourboxJPE=\null           \gdef\isjourJPE{F}
   \setbox\pagesboxJPE=\null          \gdef\ispagesJPE{F}
   \setbox\volboxJPE=\null            \gdef\isvolJPE{F}
   \setbox\preprintboxJPE=\null       \gdef\ispreprintJPE{F}
   \setbox\toappearboxJPE=\null       \gdef\istoappearJPE{F}
   \setbox\inprintboxJPE=\null        \gdef\isinprintJPE{F}
   \setbox\bookboxJPE=\null           \gdef\isbookJPE{F}  \gdef\isinbookJPE{F}
     
   \setbox\bybookboxJPE=\null         \gdef\isbybookJPE{F}
   \setbox\publisherboxJPE=\null      \gdef\ispublisherJPE{F}
}
\def\widestlabel#1{\setbox0=\hbox{#1\enspace}\refindent=\wd0\relax}
\def\ref{\refclearJPE}
\def\no#1{\gdef\isnoJPE{T}\setbox\noboxJPE=\hbox{#1}}
\def\by#1{\gdef\isbyJPE{T}\setbox\byboxJPE=\hbox{#1}}
\def\paper#1{\gdef\ispaperJPE{T}\setbox\paperboxJPE=\hbox{#1}}
\def\yr#1{\gdef\isyrJPE{T}\setbox\yrboxJPE=\hbox{#1}}
\def\jour#1{\gdef\isjourJPE{T}\setbox\jourboxJPE=\hbox{#1}}
\def\pages#1{\gdef\ispagesJPE{T}\setbox\pagesboxJPE=\hbox{#1}}
\def\vol#1{\gdef\isvolJPE{T}\setbox\volboxJPE=\hbox{\bf #1}}
\def\preprint#1{\gdef
\ispreprintJPE{T}\setbox\preprintboxJPE=\hbox{#1}}

\def\inprint{\gdef
\isinprintJPE{T}}
\def\book#1{\gdef\isbookJPE{T}\setbox\bookboxJPE=\hbox{\em #1}}
\def\publisher#1{\gdef
\ispublisherJPE{T}\setbox\publisherboxJPE=\hbox{#1}}
\def\inbook#1{\gdef\isinbookJPE{T}\setbox\bookboxJPE=\hbox{\em #1}}
\def\bybook#1{\gdef\isbybookJPE{T}\setbox\bybookboxJPE=\hbox{#1}}
\newdimen\refindent
\refindent=5em
\def\endref{\sfcode`.=1000
 \if T\isnoJPE
% \setbox0=\hbox{[\unhbox\noboxJPE\unskip]\hss\unskip\enspace}%
%   \ifdim\refindent<\wd0\relax
%      \message{\string\refno: reference is wider than
%               you pretended when using \string\widestlabel.}%
%   \fi
\hangindent\refindent\hangafter=1
      \noindent\hbox to\refindent{[\unhbox\noboxJPE\unskip]\hss}\ignorespaces
     \else  \noindent    \fi
% \if T\isnoJPE  \item{[\unhbox\noboxJPE\unskip]}
%     \else  \noindent    \fi
 \if T\isbyJPE    \unhbox\byboxJPE\unskip: \fi
 \if T\ispaperJPE \unhbox\paperboxJPE\unskip\period \fi
 \if T\isbookJPE {\it\unhbox\bookboxJPE\unskip}\if T\ispublisherJPE, \else.
\fi\fi
 \if T\isinbookJPE In {\it\unhbox\bookboxJPE\unskip}\if T\isbybookJPE,
\else\period \fi\fi
 \if T\isbybookJPE  (\unhbox\bybookboxJPE\unskip)\period \fi
 \if T\ispublisherJPE \unhbox\publisherboxJPE\unskip \if T\isjourJPE, \else\if
T\isyrJPE \  \else\period \fi\fi\fi
 \if T\istoappearJPE (To appear)\period \fi
 \if T\ispreprintJPE Pre\-print\period \fi
 \if T\isjourJPE    \unhbox\jourboxJPE\unskip\ \fi
 \if T\isvolJPE     \unhbox\volboxJPE\unskip\if T\ispagesJPE, \else\ \fi\fi
 \if T\ispagesJPE   \unhbox\pagesboxJPE\unskip\  \fi
 \if T\isyrJPE      (\unhbox\yrboxJPE\unskip)\period \fi
 \if T\isinprintJPE (in print)\period \fi
\filbreak
}
%%%%%%%%%%%%%%%%%%%%%%%%%%%%%%%%%%%%%%%%%%%%%%%%%%%%
%%%%%EOF

%% file: layout.tex
%%%%%%%%%%%%%%%%Layout
\normalbaselineskip=12pt
\baselineskip=12pt
\parskip=0pt
\parindent=22.222pt
%\beforesectionskipamount=42pt plus5pt minus2pt
\beforesectionskipamount=24pt plus0pt minus6pt
\sectionskipamount=7pt plus3pt minus0pt
\overfullrule=0pt
\hfuzz=2pt
\nopagenumbers
\headline={\ifnum\pageno>1 {\hss\tenrm-\ \folio\ -\hss} \else
{\hfill}\fi}
%%%%%%%%%%%%% SPECIAL FONTS %%%%%%%%%%%%%%%%%%%%%%%%%
%%%%%%%%workaround linux %%%%%%%%%%%%%%%%%%%%%%%%%%%%%%%%%%%%
%%%%%%%%%%%%%%%%%%%%%%%%%%%%%%%%%%%%%%%%%%%%%%
%%%%%%%%%%%%% default %%%%%%%%%%%%%%%%
\font\titlefont=ptmb at 14 pt

\font\toplinefont=cmcsc10
\font\pagenumberfont=ptmb at 10pt

%%%%%%%%%%%%%%%%%%%%%%%%%%%%%%
\newdimen\itemindent\itemindent=1.5em

\def\textindent#1{\indent\llap{#1\enspace}\ignorespaces}
\def\item{\par\noindent
\hangindent\itemindent\hangafter=1\relax
\setitemmark}
\def\setitemindent#1{\setbox0=\hbox{\ignorespaces#1\unskip\enspace}%
\itemindent=\wd0\relax
\message{|\string\setitemindent: Mark width modified to hold
         |`\string#1' plus an \string\enspace\space gap. }%
}
\def\setitemmark#1{\checkitemmark{#1}%
\hbox to\itemindent{\hss#1\enspace}\ignorespaces}
\def\checkitemmark#1{\setbox0=\hbox{\enspace#1}%
\ifdim\wd0>\itemindent
   \message{|\string\item: Your mark `\string#1' is too wide. }%
\fi}
\newcount\FOOTcount \FOOTcount=0
\def\myfoot#1{\global\advance\FOOTcount by 1\footnote{${}^{\number\FOOTcount}$}{#1}}

%% file: table.tex
% +--------------------------------------------------------------------+        
% |                                                                    |        
% |                           TABLES.TEX                               |        
% |                                                                    |        
% |                     Ray F. Cowan  15-Feb-85                        |        
% |                                                                    |        
% |                       Princeton University                         |        
% |                                                                    |        
% |                     Last Revision: 17-Apr-86                       |        
% |                                                                    |        
% |   Macros I find handy for making tables.  See TABLEDOC TEX for     |        
% |   a longer description.  The token-counting macros are straight    |        
% |   from the TeXbook's "Dirty Tricks" appendix.                      |        
% |                                                                    |        
% +--------------------------------------------------------------------+        
%                                                                               
\newbox\hdbox%                                                                  
\newcount\hdrows%                                                               
\newcount\multispancount%                                                       
\newcount\ncase%                                                                
\newcount\ncols% This is the number of primary text columns in the table.       
\newcount\nrows%                                                                
\newcount\nspan%                                                                
\newcount\ntemp%                                                                
\newdimen\hdsize%                                                               
\newdimen\newhdsize%                                                            
\newdimen\parasize%                                                             
\newdimen\spreadwidth%                                                          
\newdimen\thicksize%                                                            
\newdimen\thinsize%                                                             
\newdimen\tablewidth%                                                           
\newif\ifcentertables%                                                          
\newif\ifendsize%                                                               
\newif\iffirstrow%                                                              
\newif\iftableinfo%                                                             
\newtoks\dbt%                                                                   
\newtoks\hdtks%                                                                 
\newtoks\savetks%                                                               
\newtoks\tableLETtokens%                                                        
\newtoks\tabletokens%                                                           
\newtoks\widthspec%                                                             
%                                                                               
%  Book-keeping stuff--see how often these macros are called.                   
%                                                                               
\immediate\write15{%                                                            
CP SMSG GJMSINK TEXTABLE --> TABLE MACROS V. 851121 JOB = \jobname%             
}%                                                                              
%                                                                               
%  Turn on table diagnostics.                                                   
%                                                                               
\tableinfotrue%                                                                 
\catcode`\@=11%  Allows use of "@" in macro names, like PLAIN.TEX does.         
%  Debugging aid.  Writes #1 on the           
%                                    user's terminal and in the log file.       
%                                                                               
%  Define the \tstrut height, depth in terms of the x_height parameter.         
%                                                                               
\def\tstrut{\vrule height3.1ex depth1.2ex width0pt}%                            
\def\and{\char`\&}%  Allows us to get an `&' in the text.  This is the          
%                    same as using the PLAIN TeX macro \&.                      
\def\tablerule{\noalign{\hrule height\thinsize depth0pt}}%                      
\thicksize=1.5pt%  Default thickness for fat rules.  The user should feel       
%                  free to change this to his preference.                       
\thinsize=0.6pt%   Default thickness for thin rules.                            
\def\thickrule{\noalign{\hrule height\thicksize depth0pt}}%                     
\def\ctr#1{\hfil\ #1\hfil}%                                                     
%                                                   
%                          
%                                                                               
%  Here are things for controlling the width of the finished table.             
%                                                                               
\tablewidth=-\maxdimen%                                                         
\spreadwidth=-\maxdimen%                                                        
\def\tabskipglue{0pt plus 1fil minus 1fil}%                                     
%                                                                               
%  Stuff for centering or not.                                                  
%                                                                               
\centertablestrue%                                                              
%                                                                              
%                                                                              
%                                                                               
%  \vctr vertically centers its argument in the row.                            
%                                                                               
\parasize=4in%                                                                  
\gdef\ARGS{########}%  Produces the correct number of #'s in the preamble       
%                      by the time eveything is expanded and \halign sees       
%                      it.                                                      
\gdef\headerARGS{####}%  Same as \ARGS, but used in \header macros.             
\def\@mpersand{&}%  Allows us to get alignment tab characters later             
%                   when we have made the character "&" an active macro.        
{\catcode`\|=13%  Make |'s locally active.                                      
\gdef\letbarzero{\let|0}%  Globally define a macro that allows us to            
%                          keep active |'s from being expanded in edef's.       
\gdef\letbartab{\def|{&&}}%                                                     
\gdef\letvbbar{\let\vb|}%                                                       
%  This \def will cause active |'s read by                                      
%                            \ruledtable to be converted into double            
%                            alignment tabs.                                    
}%  End of locally active |'s.                                                  
{\catcode`\&=4%  Make these alignment tabs.                                     
\def\ampskip{&\omit\hfil&}%  This local macro skips a vertical rule.            
\catcode`\&=13%  Now make &'s into active macros.                               
\let&0%  This allows us to expand \ampskip in the next \xdef without            
%        attempting to expand the & and getting an "undefined control           
%        sequence" error.                                                       
\xdef\letampskip{\def&{\ampskip}}%                                              
\gdef\letnovbamp{\let\novb&\let\tab&}                                           
%  This will cause active &'s read by                                           
%                                   \ruledtable to be converted into            
%                                   double tabs and an \omit'ted \vrule.        
}%  End of locally active &'s.                                                  
\def\begintable{%  Here we make |'s and &'s active characters so we can         
%                  interpret them as macros.  Note that this action is          
%                  true only until we encounter the matching \endgroup          
%                  token later at the end of the \ruledtable macro.             
   \begingroup%                                                                 
   \catcode`\|=13\letbartab\letvbbar%                                           
   \catcode`\&=13\letampskip\letnovbamp%                                        
   \def\multispan##1{%  We must redefine \multispan to count the number         
%                       of primary columns, not physical columns.               
      \omit \mscount##1%                                                        
      \multiply\mscount\tw@\advance\mscount\m@ne%                               
      \loop\ifnum\mscount>\@ne \sp@n\repeat%                                    
   }%  End of \multispan macro.                                                 
   \def\|{%                                                                     
      &\omit\widevline&%                                                        
   }%                                                                           
   \ruledtable%  Now we call \ruledtable to do the real work.                   
}%  End of \begintable macro.                                                   
\long\def\ruledtable#1\endtable{%                                               
%                                                                               
%  This macro reads in the user's data entries                                  
%  and converts them into a ruled table.                                        
%                                                                               
%  Important note:  Many macros and parameters are re-defined here, and         
%  these must be kept local to the table macros to avoid conflict with          
%  their use outside of tables.  This is done by the \begingroup token          
%  macro \begintable and the \endgroup token at the end of                      
%  this macro.                                                                  
%                                                                               
   \offinterlineskip%  Needed to make rules touch each other.                   
   \tabskip 0pt%  Needed for same reason as \offinterlineskip.                  
   \def\widevline{\vrule width\thicksize}%  Make outer \vrule's wider.          
   \def\endrow{\@mpersand\omit\hfil\crnorm\@mpersand}%                          
   \def\crthick{\@mpersand\crnorm\thickrule\@mpersand}%                         
   \def\crthickneg##1{\@mpersand\crnorm\thickrule                               
          \noalign{{\skip0=##1\vskip-\skip0}}\@mpersand}%                       
   \def\crnorule{\@mpersand\crnorm\@mpersand}%                                  
   \def\crnoruleneg##1{\@mpersand\crnorm                                        
          \noalign{{\skip0=##1\vskip-\skip0}}\@mpersand}%                       
   \let\nr=\crnorule%  A shorter abbreviation.                                  
   \def\endtable{\@mpersand\crnorm\thickrule}%                                  
   \let\crnorm=\cr%  Allows us to use \cr for our own purposes.                 
%                                                                               
%  Cause user-typed \cr's to follow a row with a \tablerule.                    
%                                                                               
   \edef\cr{\@mpersand\crnorm\tablerule\@mpersand}%                             
   \def\crneg##1{\@mpersand\crnorm\tablerule                                    
          \noalign{{\skip0=##1\vskip-\skip0}}\@mpersand}%                       
   \let\ctneg=\crthickneg                                                       
   \let\nrneg=\crnoruleneg                                                      
   \the\tableLETtokens%  Get the user's extra \let's, if any.                   
%                                                                               
%  Put the data entries into a token register so we can scan through them       
%  and see what the user is asking us to do.                                    
%                                                                               
   \tabletokens={&#1}%  We add an extra alignment tab to the beginning          
%                       of the first row to allow for the first \vrule.         
%                                                                               
%  Now count how many rows are in the table and return the result in            
%  count register \nrows; do the same for columns, and return that              
%  in register \ncols.                                                          
%                                                                               
   \countROWS\tabletokens\into\nrows%                                           
   \countCOLS\tabletokens\into\ncols%                                           
%                                                                               
%  Now do a little arithmetic to convert the number of primary columns          
%  into the number of physical columns that the alignment preamble must         
%  prepare for;  similarly for rows.                                            
%                                                                               
   \advance\ncols by -1%                                                        
   \divide\ncols by 2%                                                          
   \advance\nrows by 1%                                                         
%                                                                               
%  Tell the user how many rows and columns we found in his data, if he          
%  wants to know.                                                               
%                                                                               
   \iftableinfo %                                                               
      \immediate\write16{[Nrows=\the\nrows, Ncols=\the\ncols]}%                 
   \fi%                                                                         
%                                                                               
%  Now we actually go ahead and produce the table.                              
%                                                                               
   \ifcentertables                                                              
      \ifhmode \par\fi%  Make sure we are in vertical mode.                     
      \line{%  The final table comes out as an \hbox of width the \hsize.       
      \hss%  The final table will be centered left-to-right.                    
   \else %                                                                      
      \hbox{%                                                                   
   \fi                                                                          
      \vbox{%                                                                   
         \makePREAMBLE{\the\ncols}%  Generate the preamble.                     
         \edef\next{\preamble}%  This line and the next line force the          
         \let\preamble=\next%    expansion of all \ARGS tokens into the         
%                                appropriate number of #'s.                     
         \makeTABLE{\preamble}{\tabletokens}%  Go do the \halign here.          
      }%  End of \vbox.                                                         
      \ifcentertables \hss}\else }\fi%  Finish the centering effect.            
%                                       It is important that no spaces          
%                                       follow the two `}' here.                
%  }%  End of \line.                                                            
   \endgroup%  Return all local macros and parameters to their outside          
%              values.                                                          
   \tablewidth=-\maxdimen%  Reset \tablewidth to normal.                        
   \spreadwidth=-\maxdimen% Same for \spreadwidth.                              
}%  End of macro \ruledtable.                                                   
\def\makeTABLE#1#2{%  Does an \halign for the \ruledtable macro.                
   {%  Start of local parameter values.                                         
   \let\ifmath0%     These macros would cause trouble if they were to be        
   \let\header0%     expanded in the following \xdef; we \let them be           
   \let\multispan0%  equal to a digit, because digits can't be expanded.        
%                                                                               
%  Set up the width specification here.                                         
%                                                                               
   \ncase=0%                                                                    
   \ifdim\tablewidth>-\maxdimen \ncase=1\fi%                                    
   \ifdim\spreadwidth>-\maxdimen \ncase=2\fi%                                   
   \relax%  This \relax is absolutely necessary, without it the following       
%           \ifcase will always take \ncase=0.                                  
%                                                                               
   \ifcase\ncase %                                                              
      \widthspec={}%                                                            
   \or %                                                                        
      \widthspec=\expandafter{\expandafter t\expandafter o%                     
                 \the\tablewidth}%                                              
   \else %                                                                      
      \widthspec=\expandafter{\expandafter s\expandafter p\expandafter r%       
                 \expandafter e\expandafter a\expandafter d%                    
                 \the\spreadwidth}%                                             
   \fi %                                                                        
%\out{Widthspec=[\the\widthspec]}%                                              
%\out{Preamble=[\preamble]}%                                                    
   \xdef\next{%  We must force the preamble to be expanded BEFORE the           
      \halign\the\widthspec{%                                                   
%        \halign is done;  this \edef\next{...}\next construction               
%                does the trick.                                                
      #1%  This is the preamble text.                                           
      \noalign{\hrule height\thicksize depth0pt}%  Makes the top \hrule.        
      \the#2\endtable%  This is the main body.                                  
%                                                                               
%     \noalign{\hrule height0.7pt depth0pt}%  Makes the last \hrule.            
      }%  End of \halign.                                                       
   }%  End of \next.                                                            
   }%  End of local values.                                                     
   \next%  This \next must be outside of the local values, because now          
%          we want those troublesome macros in the \let's above to have         
%          their normal actions.                                                
}%  End of macro \makeTABLE.                                                    
\def\makePREAMBLE#1{%  This macro generates the necessary preamble for a        
%                      ruled table with #1 primary columns.                     
%                      (Primary columns means the number of columns NOT         
%                       counting those used for vertical rules.)                
   \ncols=#1%  Get the number of columns desired.                               
   \begingroup%  Start local parameter definitions.                             
   \let\ARGS=0%  This is the key to the whole thing; it prevents \ARGS          
%                from being expanded in the following \edef's.                  
   \edef\xtp{\widevline\ARGS\tabskip\tabskipglue%                               
   &\ctr{\ARGS}\tstrut}%  A 1-column preamble.  Gets the sizing right.          
   \advance\ncols by -1%  One column has been generated; decrement the          
%                         counter.                                              
   \loop%  Append as many further columns as needed to the preamble.            
      \ifnum\ncols>0 %                                                          
      \advance\ncols by -1%                                                     
      \edef\xtp{\xtp&\vrule width\thinsize\ARGS&\ctr{\ARGS}}%                   
   \repeat                                                                      
   \xdef\preamble{\xtp&\widevline\ARGS\tabskip0pt%                              
   \crnorm}%  Adds the last \vrule.                                             
   \endgroup%  End of local parameters.                                         
}%  End of macro \makePREAMBLE.                                                 
\def\countROWS#1\into#2{%  This counts the number of rows in #1 by              
%                          looking for control sequences that end a row,        
%                          e.g., \cr, \crthick, etc., and puts the result       
%                          into count register #2.                              
   \let\countREGISTER=#2%                                                       
   \countREGISTER=0%                                                            
%  \out{In countROWS:  tokens are [\the#1]}%                                    
   \expandafter\ROWcount\the#1\endcount%                                        
}%                                                                              
\def\ROWcount{%                                                                 
   \afterassignment\subROWcount\let\next= %                                     
}%                                                                              
\def\subROWcount{%                                                              
%  \out{In subROWcount:  next is [\meaning\next]}%  Debugging aid.              
   \ifx\next\endcount %                                                         
      \let\next=\relax%                                                         
   \else%                                                                       
      \ncase=0%                                                                 
      \ifx\next\cr %                                                            
         \global\advance\countREGISTER by 1%                                    
         \ncase=0%                                                              
      \fi%                                                                      
      \ifx\next\endrow %                                                        
         \global\advance\countREGISTER by 1%                                    
         \ncase=0%                                                              
      \fi%                                                                      
      \ifx\next\crthick %                                                       
         \global\advance\countREGISTER by 1%                                    
         \ncase=0%                                                              
      \fi%                                                                      
      \ifx\next\crnorule %                                                      
         \global\advance\countREGISTER by 1%                                    
         \ncase=0%                                                              
      \fi%                                                                      
      \ifx\next\crthickneg %                                                    
         \global\advance\countREGISTER by 1%                                    
         \ncase=0%                                                              
      \fi%                                                                      
      \ifx\next\crnoruleneg %                                                   
         \global\advance\countREGISTER by 1%                                    
         \ncase=0%                                                              
      \fi%                                                                      
      \ifx\next\crneg %                                                         
         \global\advance\countREGISTER by 1%                                    
         \ncase=0%                                                              
      \fi%                                                                      
      \ifx\next\header %                                                        
%     \out{In subROWcount:  next=header, ncase set=1}%                          
         \ncase=1%                                                              
      \fi%                                                                      
%     \out{In subROWcount:  ncase is [\the\ncase]}%                             
      \relax%                                                                   
      \ifcase\ncase %                                                           
         \let\next\ROWcount%                                                    
%        \out{subROWcount---> ncase=\the\ncase}%                                
      \or %                                                                     
         \let\next\argROWskip%                                                  
%        \out{subROWcount---> ncase=\the\ncase}%                                
      \else %                                                                   
      \fi%                                                                      
   \fi%                                                                         
%  \out{subROWcount---> NEXT=\meaning\next}%                                    
   \next%                                                                       
}%  End of macro \subROWcount.                                                  
\def\counthdROWS#1\into#2{%                                                     
\dvr{10}%                                                                       
   \let\countREGISTER=#2%                                                       
   \countREGISTER=0%                                                            
\dvr{11}%                                                                       
%  \out{In counthdROWS:  tokens are [\the#1]}%                                  
\dvr{13}%                                                                       
   \expandafter\hdROWcount\the#1\endcount%                                      
\dvr{12}%                                                                       
}%                                                                              
\def\hdROWcount{%                                                               
   \afterassignment\subhdROWcount\let\next= %                                   
}%                                                                              
\def\subhdROWcount{%                                                            
%\out{In subhdROWcount:  next is [\meaning\next]}%                              
   \ifx\next\endcount %                                                         
      \let\next=\relax%                                                         
   \else%                                                                       
      \ncase=0%                                                                 
      \ifx\next\cr %                                                            
         \global\advance\countREGISTER by 1%                                    
         \ncase=0%                                                              
      \fi%                                                                      
      \ifx\next\endrow %                                                        
         \global\advance\countREGISTER by 1%                                    
         \ncase=0%                                                              
      \fi%                                                                      
      \ifx\next\crthick %                                                       
         \global\advance\countREGISTER by 1%                                    
         \ncase=0%                                                              
      \fi%                                                                      
      \ifx\next\crnorule %                                                      
         \global\advance\countREGISTER by 1%                                    
         \ncase=0%                                                              
      \fi%                                                                      
      \ifx\next\header %                                                        
%\out{In subhdROWcount:  next=header, ncase set=1}%                             
         \ncase=1%                                                              
      \fi%                                                                      
%\out{In subhdROWcount:  ncase is [\the\ncase]}%                                
\relax%                                                                         
      \ifcase\ncase %                                                           
         \let\next\hdROWcount%                                                  
%\out{subhdROWcount---> ncase=\the\ncase}%                                      
      \or%                                                                      
         \let\next\arghdROWskip%                                                
%\out{subhdROWcount---> ncase=\the\ncase}%                                      
      \else %                                                                   
      \fi%                                                                      
   \fi%                                                                         
%\out{subhdROWcount---> NEXT=\meaning\next}%                                    
   \next%                                                                       
}%                                                                              
{\catcode`\|=13\letbartab                                                       
\gdef\countCOLS#1\into#2{%                                                      
%  \out{In countCOLS:  tokens are [\the#1]}                                     
   \let\countREGISTER=#2%                                                       
   \global\countREGISTER=0%                                                     
   \global\multispancount=0%                                                    
   \global\firstrowtrue                                                         
   \expandafter\COLcount\the#1\endcount%                                        
   \global\advance\countREGISTER by 3%                                          
   \global\advance\countREGISTER by -\multispancount                            
%  \out{countCOLS-->[\the\countREGISTER]}                                       
}%                                                                              
\gdef\COLcount{%                                                                
   \afterassignment\subCOLcount\let\next= %                                     
}%                                                                              
{\catcode`\&=13%                                                                
\gdef\subCOLcount{%                                                             
%\out{In subCOLcount: next is [\meaning\next]}                                  
   \ifx\next\endcount %                                                         
      \let\next=\relax%                                                         
   \else%                                                                       
      \ncase=0%                                                                 
      \iffirstrow                                                               
         \ifx\next& %                                                           
            \global\advance\countREGISTER by 2%                                 
            \ncase=0%                                                           
         \fi%                                                                   
         \ifx\next\span %                                                       
            \global\advance\countREGISTER by 1%                                 
            \ncase=0%                                                           
         \fi%                                                                   
         \ifx\next| %                                                           
            \global\advance\countREGISTER by 2%                                 
            \ncase=0%                                                           
         \fi                                                                    
         \ifx\next\|                                                            
            \global\advance\countREGISTER by 2%                                 
            \ncase=0%                                                           
         \fi                                                                    
         \ifx\next\multispan                                                    
            \ncase=1%                                                           
            \global\advance\multispancount by 1%                                
         \fi                                                                    
         \ifx\next\header                                                       
            \ncase=2%                                                           
         \fi                                                                    
         \ifx\next\cr       \global\firstrowfalse \fi                           
         \ifx\next\endrow   \global\firstrowfalse \fi                           
         \ifx\next\crthick  \global\firstrowfalse \fi                           
         \ifx\next\crnorule \global\firstrowfalse \fi                           
         \ifx\next\crnoruleneg \global\firstrowfalse \fi                        
         \ifx\next\crthickneg  \global\firstrowfalse \fi                        
         \ifx\next\crneg       \global\firstrowfalse \fi                        
      \fi%  End of \iffirstrow.                                                 
\relax%\out{subCOL-->  ncase=[\the\ncase]}                                      
% \out{subCOL-->  next=\meaning\next}                                           
      \ifcase\ncase %                                                           
         \let\next\COLcount%                                                    
      \or %                                                                     
         \let\next\spancount%                                                   
      \or %                                                                     
         \let\next\argCOLskip%                                                  
      \else %                                                                   
      \fi %                                                                     
   \fi%                                                                         
%  \out{subCOL-->  countREGISTER=[\the\countREGISTER]}                          
   \next%                                                                       
}%                                                                              
\gdef\argROWskip#1{%                                                            
%  Deletes the next balanced, undelimited argument from a                       
%                 token list.                                                   
% \out{---> Entering argROWskip <---}                                           
% \out{In argROWskip:  deleted arg is [#1]}%                                    
   \let\next\ROWcount \next%                                                    
}%  End of macro \argskip.                                                      
\gdef\arghdROWskip#1{%                                                          
%  Deletes the next balanced, undelimited argument from a                       
%                 token list.                                                   
% \out{---> Entering arghdROWskip <---}                                         
% \out{In arghdROWskip:  deleted arg is [#1]}%                                  
   \let\next\ROWcount \next%                                                    
}%  End of macro \arghdROWskip.                                                 
\gdef\argCOLskip#1{%                                                            
%  Deletes the next balanced, undelimited argument from a                       
%                 token list.                                                   
% \out{---> Entering argCOLskip <---}                                           
% \out{In argCOLskip:  deleted arg is [#1]}%                                    
   \let\next\COLcount \next%                                                    
}%  End of macro \argskip.                                                      
}%  End of active &'s.                                                          
}%  End of active |'s.                                                          
\def\spancount#1{%\out{spancount--->\meaning#1}                                 
   \nspan=#1\multiply\nspan by 2\advance\nspan by -1%                           
   \global\advance \countREGISTER by \nspan                                     
%  \out{number spancount--->\the\nspan; \the\countREGISTER}                     
   \let\next\COLcount \next}%                                                   
\def\dvr#1{\relax}%                                                             
% \omit\hfil%                                                                   
% \parindent=0pt\hsize=1.1in\valign{%                                           
% \vfil#\vfil&\vfil#\vfil\cr\hfil\hbox{\ Added to\ }\hfil&%                     
% \hfil\hbox{\ empty events\ }\hfil\cr}\hfil%                                   
\def\header#1{%                                                                 
\dvr{1}{\let\cr=\@mpersand%                                                     
\hdtks={#1}%                                                                    
%\out{In header:  hdtks=[\the\hdtks]}%                                          
\counthdROWS\hdtks\into\hdrows%                                                 
\advance\hdrows by 1%                                                           
\ifnum\hdrows=0 \hdrows=1 \fi%                                                  
%\out{In header:  Nhdrows=[\the\hdrows]}%                                       
\dvr{5}\makehdPREAMBLE{\the\hdrows}%                                            
%\out{In header:  headerpreamble=[\headerpreamble]}%                            
\dvr{6}\getHDdimen{#1}%                                                         
%\out{In header:  hdsize=[\the\hdsize]}%                                        
%\striplastCR{#1}%                                                              
{\parindent=0pt\hsize=\hdsize{\let\ifmath0%                                     
\xdef\next{\valign{\headerpreamble #1\crnorm}}}\dvr{7}\next\dvr{8}%             
}%                                                                              
}\dvr{2}}%  End of macro \header.                                               
\def\makehdPREAMBLE#1{%This macro generates the necessary preamble for a        
\dvr{3}%                                                                        
%                      ruled table with \ncols primary columns.                 
%                      (Primary columns means the number of columns NOT         
%                       counting those used for vertical rules.                 
\hdrows=#1%  Get the number of columns desired.                                 
{%  Start local parameter definitions.                                          
\let\headerARGS=0%                                                              
%  This is the key to the whole thing; it prevents \ARGS                        
\let\cr=\crnorm%                                                                
%                from being expanded in the followin \edef's.                   
\edef\xtp{\vfil\hfil\hbox{\headerARGS}\hfil\vfil}%                              
\advance\hdrows by -1%  One row has been generated; decrement the               
%                         counter.                                              
\loop%  Append as many further rows as needed to the preamble.                  
\ifnum\hdrows>0%                                                                
\advance\hdrows by -1%                                                          
\edef\xtp{\xtp&\vfil\hfil\hbox{\headerARGS}\hfil\vfil}%                         
\repeat%                                                                        
\xdef\headerpreamble{\xtp\crcr}%                                                
}%  End of local parameters.                                                    
\dvr{4}}%  End of \makehdPREAMBLE.                                              
\def\getHDdimen#1{%                                                             
%\out{In getHDdimen:  Arg 1=[#1]}%                                              
\hdsize=0pt%                                                                    
\getsize#1\cr\end\cr%                                                           
}%  End of macro getHDdimen.                                                    
\def\getsize#1\cr{%                                                             
%\out{In getsize:  Arg 1=[#1]}%                                                 
%  Here we have to check arg#1 and see if the first token in #1 is an           
%    \end; if so, we stop, else we check the width of arg#1.                    
%  We recall that each arg#1 will be terminated with a \cr token.               
\endsizefalse\savetks={#1}%                                                     
%\out{In getsize:  the savetks = [\the\savetks]}%                               
\expandafter\lookend\the\savetks\cr%                                            
%\out{In getsize:  ifendsize = [\meaning\ifendsize]}%                           
\relax \ifendsize \let\next\relax \else%                                        
\setbox\hdbox=\hbox{#1}\newhdsize=1.0\wd\hdbox%                                 
\ifdim\newhdsize>\hdsize \hdsize=\newhdsize \fi%                                
%\out{In getsize:  hdsize=[\the\hdsize]}%                                       
%\out{In getsize:  newhdsize=[\the\newhdsize]}%                                 
\let\next\getsize \fi%                                                          
\next%                                                                          
}%                                                                              
\def\lookend{\afterassignment\sublookend\let\looknext= }%                       
\def\sublookend{\relax%                                                         
%\out{In sublookend:  looknext = [\looknext]}%                                  
\ifx\looknext\cr %                                                              
%\out{In sublooknext:  looknext=cr}%                                            
\let\looknext\relax \else %                                                     
%\out{In sublooknext:  looknext/=cr}%                                           
   \relax                                                                       
   \ifx\looknext\end \global\endsizetrue \fi%                                   
   \let\looknext=\lookend%                                                      
    \fi \looknext%                                                              
}%                                                                              
%                                                                               
%  Allow the user to make his own names for crthick, etc.                       
%                                                                               
\def\tablelet#1{%                                                               
   \tableLETtokens=\expandafter{\the\tableLETtokens #1}%                        
}%                                                                              
\catcode`\@=12%  Change @'s back to their normal category code.                 